\documentclass[%
 reprint,
 amsmath,amssymb,
 aps,
]{revtex4-2}

\usepackage{graphicx}% Include figure files
\usepackage{dcolumn}% Align table columns on decimal point
\usepackage{bm}
\usepackage{romannum}

\usepackage{amsmath}
\usepackage{varwidth}
\usepackage{graphicx}
\usepackage{xcolor}
\usepackage{multirow}
\usepackage[caption=false]{subfig}
\usepackage{booktabs}   
\usepackage{relsize}

\begin{document}

\preprint{APS/123-QED}

\title{Dwarf Galaxies United by Dark Bosons}% Force line breaks with \\
\author{Alvaro Pozo$^{1,2}$, Tom Broadhurst$^{1,2,3}$, George~F.~ Smoot$^{1,4,5,6,7}$, Tzihong Chiueh$^{8,9}$,  Hoang Nhan Luu$^{1}$, Mark Vogelsberger$^{10}$, Philip Mocz$^{11,12}$\\ }

\affiliation{$^{1}$\textit{Donostia International Physics Center (DIPC), Basque Country UPV/EHU, E-48080 San Sebastian, Spain;\\ email:alvaro.pozolarrocha@bizkaia.eu; tom.j.broadhurst@gmail.com;}\\
$^{2}$\textit{Department of Theoretical Physics, University of the Basque Country UPV/EHU, E-48080 Bilbao, Spain}\\
$^{3}$\textit{Ikerbasque, Basque Foundation for Science, E-48011 Bilbao, Spain}\\
$^{4}$\textit{Institute for Advanced Study and Department of Physics, IAS TT \& WF Chao Foundation Professor, Hong Kong University of Science and Technology, Hong Kong}\\
$^{5}$\textit{Energetic Cosmos Laboratory, Nazarbayev University, Nursultan, Kazakhstan}\\
$^{6}$\textit{Physics Department, University of California at  Berkeley CA 94720 Emeritus}\\
$^{7}$\textit{Paris Centre for Cosmological Physics, APC, AstroParticule et Cosmologie, Universit\'{e} de Paris, CNRS/IN2P3, CEA/lrfu,Universit\'{e} Sorbonne Paris Cit\'{e}, 10, rue Alice Domon et Leonie Duquet,	75205 Paris CEDEX 13, France  Emeritus}\\
$^{8}$\textit{Department of Physics, National Taiwan University, Taipei 10617, Taiwan}\\
$^{9}$\textit{National Center for Theoretical Sciences, National Taiwan University, Taipei 10617, Taiwan}\\
$^{10}$\textit{Department of Physics, Kavli Institute for Astrophysics and Space Research, Massachusetts Institute of Technology, Cambridge, MA 02139, USA}
$^{11}$\textit{Department of Astrophysical Sciences, Princeton University, 4 Ivy Lane, Princeton, NJ, 08544, USA}\\
$^{12}$\textit{Lawrence Livermore National Laboratory, 7000 East Ave, Livermore, CA 94550, USA}}

\date{\today}% It is always \today, today,

\begin{abstract}
\textbf{Low mass galaxies in the Local Group are dominated by dark matter and comprise the well studied ``dwarf Spheroidal" (dSph) class, with typical masses of $10^{9-10}M_\odot$ and also the equally numerous ``ultra faint dwarfs" (UFD), discovered recently, that are distinctly smaller and denser with masses of only $10^{7-8}M_\odot$. This bimodality amongst low mass galaxies contrasts with the scale free continuity expected for galaxies formed under gravity, as in the standard Cold Dark Matter (CDM) model for heavy particles. Within each dwarf class we find the core radius $R_c$ is inversely related to velocity dispersion $\sigma$, quite the opposite of standard expectations, but indicative of dark matter in a Bose-Einstein state, where the Uncertainty Principle requires $R_c \times \sigma$ is fixed by Plancks constant, $h$. The corresponding boson mass, $m_b=h/R_c \sigma$, differs by one order of magnitude between the UFD and dSph classes, with $10^{-21.4}$eV and $10^{-20.3}$eV respectively. Two boson species is reinforced by parallel relations seen between the central density and radius of UFD and dSph dwarfs respectively, each matching the steep prediction, $\rho_c \propto R_c^{-4}$, for soliton cores in the ground state. Furthermore, soliton cores accurately fit the stellar profiles of UFD and dSph dwarfs where prominent, dense cores appear surrounded by low density halos, as predicted by our simulations. Multiple bosons may point to a String Theory interpretation for dark matter, where a discrete mass spectrum of axions is generically predicted to span many decades in mass, offering a unifying "Axiverse" interpretation for the observed "diversity" of dark matter dominated dwarf galaxies.} \\

\end{abstract}
%implying

\maketitle
Dark Matter is commonly understood to be non-relativistic, a characteristic necessary for its gravitational role in galaxy formation and explaining the spectrum of Cosmic Microwave Background (CMB) fluctuations \cite{Planck:2018}. However, the conventional interpretation involving heavy particles faces challenges, such as the notable absence of new particle signatures in laboratory experiments \cite{Aprile:2018,CMS:2020}. Furthermore, inconsistencies arise between the predictions of Cold Dark Matter (CDM) and the peculiar properties observed in dwarf galaxies \cite{Moore:1994,deBlok:2010,Marsh:2015,Klypin:1999, Safarzadeh:2021}. Some of these issues are mitigated by addressing the missing satellite problem, but even Ultra Faint Dwarf (UFD) galaxies remain in tension \cite{Kim:2018}. Alternatively, the concept of dark matter as an inherently non-relativistic Bose Einstein condensate \cite{Widrow:1993,Hu:2000} has gained attention through initial simulations. These simulations reveal pervasive interference on the de Broglie wavelength within galaxies and filaments, coining the term 'Wave Dark Matter,' or $\psi$DM \cite{Schive:2014,Hui:2020}. In this model, bosons cannot be confined to scales smaller than the de Broglie scale, leading to the suppression of dwarf galaxy formation and the emergence of a prominent soliton core \cite{Schive:2014,Schive:20142,Mocz:2017,Veltmaat:2018,Niemeyer:2020} in every galaxy. Here, self-gravity balances the effective pressure from the Uncertainty Principle in the ground state. Crucially, smaller galaxies are predicted to have wider cores and lower density because the soliton is larger at lower momentum, a hypothesis we explore in this study.

%Dark Matter is widely understood to be non-relativistic, i.e. cold, as required to form galaxies gravitationally and for explaining the spectrum of CMB fluctuation\cite{Planck:2018}. However, the standard heavy particle interpretation faces a stringent laboratory absence of new particle signatures\cite{Aprile:2018,CMS:2020} and several inconsistencies between CDM predictions and the puzzling properties of dwarf galaxies\cite{Moore:1994,deBlok:2010,Marsh:2015,Klypin:1999, Safarzadeh:2021} with some of them partially alleviated as the missing satellite problem, even UFD kind galaxies are still in tension \cite{Kim:2018}. Alternatively, the inherently non-relativistic possibility of dark matter as a Bose Einstein condensate \cite{Widrow:1993,Hu:2000} has been awakened by the first simulations in this context, revealing that pervasive interference on the de Broglie wavelength is predicted within galaxies and filaments in this context and termed "Wave Dark Matter", $\psi$DM \cite{Schive:2014,Hui:2020}. Bosons cannot be confined to less than the de Broglie scale thereby suppressing dwarf galaxy formation and generating a prominent soliton core \cite{Schive:2014,Schive:20142,Mocz:2017,Veltmaat:2018,Niemeyer:2020} in every galaxy, where self gravity balances the effective pressure from the Uncertainty Principle in the ground state. Crucially smaller galaxies are predicted to have wider cores of and lower density because the soliton is larger at lower momentum, which we test here.

  \begin{figure*}[htp]
	\centering
    \captionsetup{format=plain}
    \includegraphics[scale=0.58]{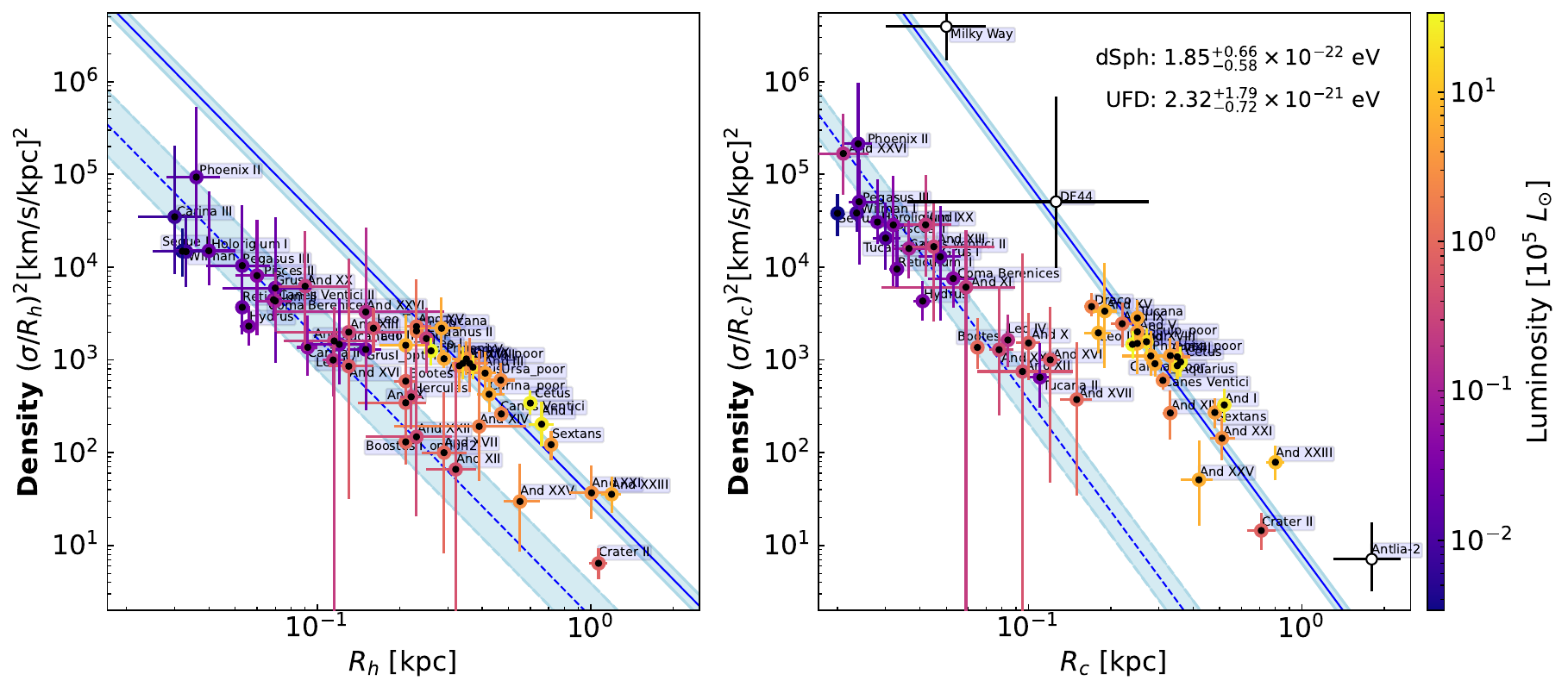}
    \caption{{\bf Left  panel: Density vs. Half-light radius}. Here, we plot the central density, $(\sigma/R_h)^2$, for each local dwarf (named on the plot), as reported by various groups (refer to the Supplement). These are color-coded by luminosity, revealing a clear distinction between the UFD and dSph classes. The data forms two parallel power-law fits shown in blue.  {\bf Right panel: Density vs. core radius.}  Here, we depict the density within our fitted core radius for each dwarf, $(\sigma/R_c)^2$, utilizing the soliton form for the core (refer to the Supplement for all individual dwarf fits to $\psi$DM and Plummer profiles). This approach results in sharper parallel relationships between the UFD and dSph dwarfs, showing a good fit to the slope $d\log \rho_c/d\log R_c = -4$. This slope corresponds to the time-independent soliton solution of the Schrodinger-Poisson relation, where a higher soliton mass leads to a narrower core. The slopes of the blue lines are constrained to be -4, as predicted by wave dark matter. For the left panel, the slope is -3 due to the $R_{\text{half}}/R_{c}$ relation. Core densities reported for the Milky Way, DF44, and Antlia-2 have been added and are seen to be consistent with the lighter boson, aligning with the dSph class. We have included core densities reported for the Milky Way, DF44, and Antlia-2, found to be consistent with the lighter boson, aligning with the dSph class. It is important to note that $R_c$ cannot be as efficiently constrained for these cases, and thus they have been added in an illustrative manner (without accounting for the calculations), colored in white. Their values are presented without ensuring their reasonability as with the other cases.}\label{figrcl}

\end{figure*}

 \begin{figure*}[htp]
	\centering
    \includegraphics[scale=0.8]{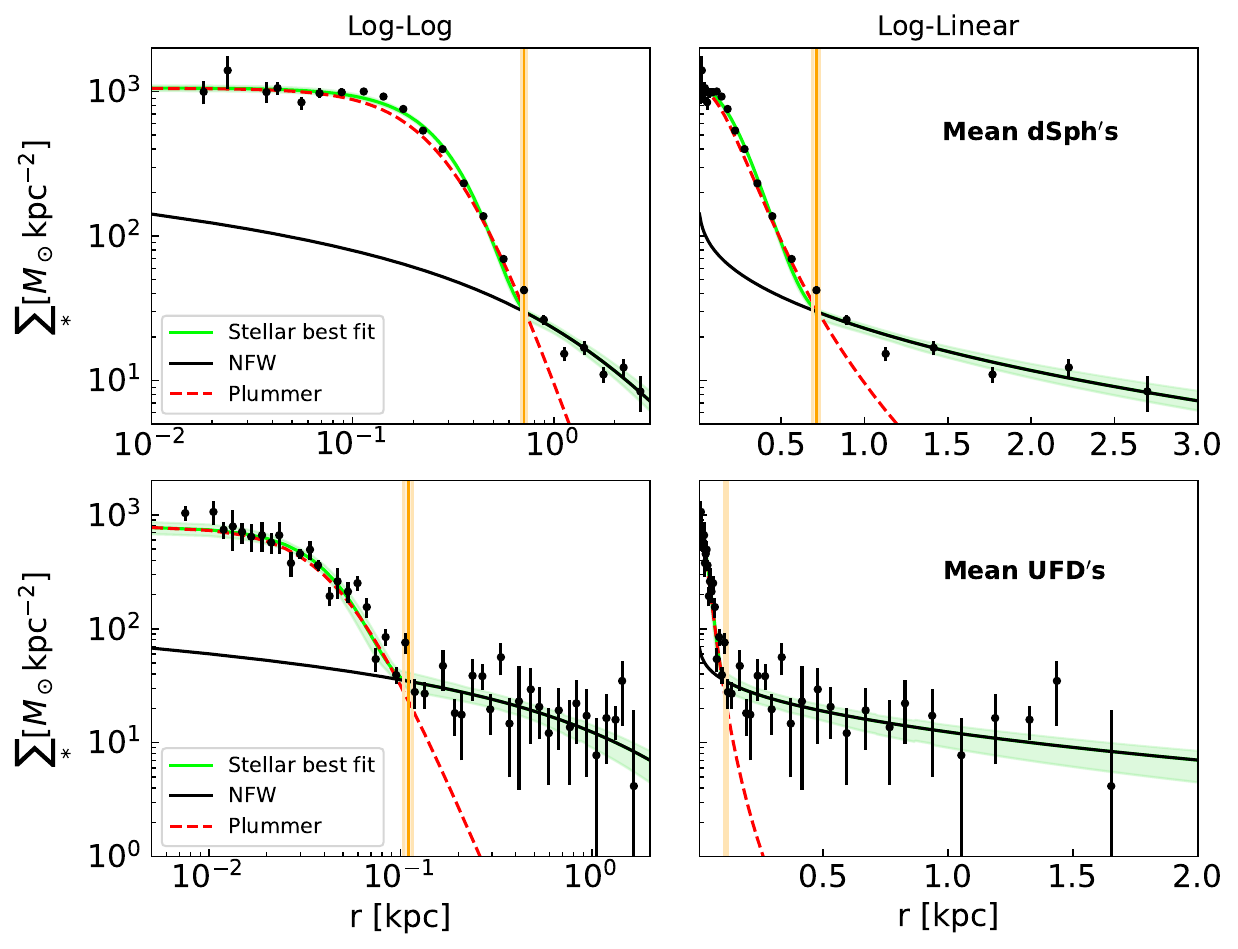}
	\caption{{\bf Top  panel: Dwarf Spheroidal Galaxies.} The mean star count profile, scaled to the mean core radius of all dSph dwarfs listed in Table \ref{tabla:1}, reveals prominent cores in both dSph dwarfs relative to the halo, which extends to several times the core radius. While a standard Plummer profile (red dashed curve) roughly fits the core region, it falls significantly short at larger radii. In contrast, the $\psi$DM profile, with its inherent core-halo structure, accurately fits the core from the soliton component and extends to the halo when azimuthally averaged over the excited states approximating the NFW form. The soliton profile has only one free parameter, the boson mass, $m_b$, determining the scale radius of the soliton. The sharp density drop between the core and the halo, by a factor of $\simeq 30$, is a characteristic feature of $\psi$DM at a transition radius marked by the vertical orange band. The best-fit MCMC profile parameters are tabulated in Table \ref{tabla:collage}. {\bf Lower panel: Ultra Faint Dwarfs.} The mean profile, averaged over all resolved profiles of Ultra Faint dwarfs listed in Table \ref{tabla:2}, exhibits the predicted $\psi$DM core-halo structure. This includes a marked transition in density between the core (highlighted in orange), with the best-fit MCMC profile parameters tabulated in Table \ref{tabla:collage}. }\label{figmeans}

\end{figure*} 

\begin{figure*}[htp]
%\begin{varwidth}{\linewidth}
	\centering
    \includegraphics[scale=0.75]{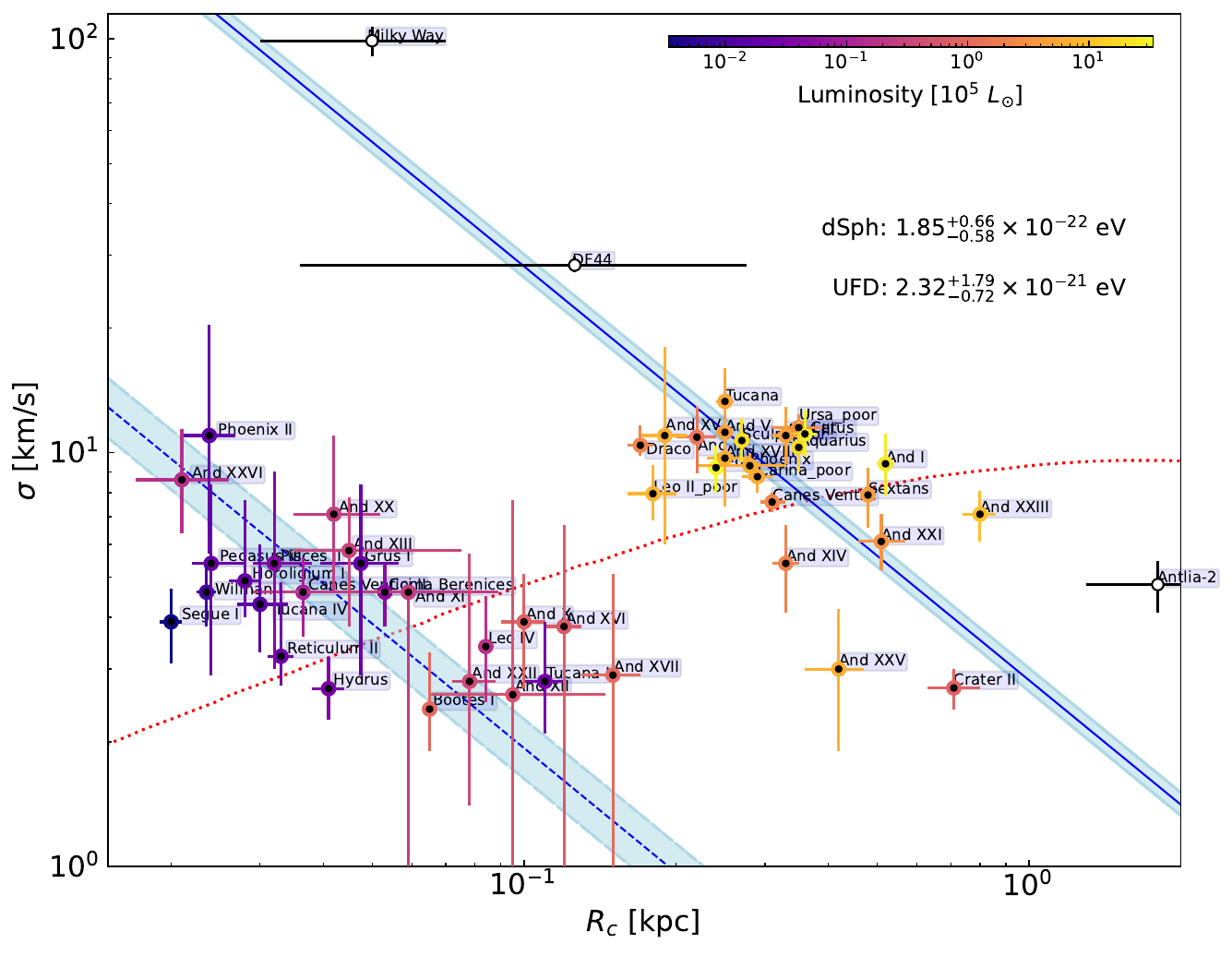}
    \caption{{\bf Velocity dispersion vs. core radius}. Here, we plot the observed velocity dispersion against the core radius for all dSph and UFD dwarfs, comparing them with the inverse relation required by the Uncertainty Principle. The best-fit blue lines to the UFD and dSPh dwarfs separately are indicated, with the corresponding boson masses of Wave-DM derived from the normalization shown in the legend. The slopes of the blue lines are constrained to be -1, as predicted by wave dark matter. Additionally, the CDM-related prediction \citep{Walker:2009} is shown as a thin red curve, where galaxies with NFW profiles are larger with increasing mass, contrary to the behavior expected in $\psi$DM.  }\label{figrcd}
% \end{varwidth}  
\end{figure*} 

Here, we rigorously test the distinctive soliton predictions by examining well-resolved dwarf galaxies within the Local Group that orbit both Andromeda and the Milky Way. We utilize their star count profiles and velocity dispersion profiles (refer to the Supplement for details). Initially, we plot the reported half-light radius, $R_{h}$, against the standard dynamical measure of density within this radius, $M(<R_h) \propto \sigma^2 R_{h}$. This allows the central density to scale to an order unity dimensionless constant $\alpha$, expressed as $4\pi G\rho_{h}=\alpha\sigma^2/R_{h}^2$. The left panel of Figure \ref{figrcl} displays this correlation, color-coded by stellar luminosity, revealing two steep parallel relations. UFD galaxies follow a relatively small and dense track compared to the dSph dwarfs. Both classes of dwarfs exhibit a similar, unexpectedly negative correlation, trending towards lower density and larger radius, as illustrated in Figure \ref{figrcl}.

%Here we test these unique soliton predictions, examining all well resolved Local Group dwarf galaxies orbiting Andromeda and the Milky Way, using their star count profiles and velocity dispersion profiles (See Supplement). We first plot the reported the half-light radius, $R_{h}$ against the standard dynamical measure of density, within this radius, $M(<R_h) \propto \sigma^2 R_{h}$, so the central density scales to order unity dimensionless constant $\alpha$, as $4\pi G\rho_{h}=\alpha\sigma^2/R_{h}^2$. This is plotted in Figure \ref{figrcl} (left panel) and colour coded by stellar luminosity, revealing two steep parallel relations with UFD galaxies following a relatively small and dense track compared to the dSph dwarfs. Both classes of dwarf show a similar, surprisingly negative correlation, towards lower density and larger radius for UFD and dSph dwarfs shown in Figure \ref{figrcl}.

 This study is built upon the premise that stars might effectively trace dark matter, a notion previously deemed controversial. Numerous distinguishing features among different dark matter models not only support this idea but also illustrate how stellar profiles should evolve uniquely in each model. In the context of cold dark matter, the scale-free formation of dark matter structures extends to extremely small scales, encompassing lower halo masses with higher DM concentrations. Conversely, $\psi$DM inherently exhibits small-scale suppression in the power spectrum, restricting the minimum scale of structure below the de Broglie length determined by the boson mass. Consequently, structures below approximately $10^9 M_{\odot}$ are suppressed for a boson mass of $10^{-22}$ eV, aligned with the observed $\simeq 0.3$ kpc scale of dwarf galaxy cores. This limitation leads to a delayed galaxy formation in $\psi$DM compared to CDM, resulting in a distinct evolutionary history. In CDM, the formation of the first ``filamentary" structures, consisting of low-mass subhaloes, occurs earlier. Conversely, in $\psi$DM, there is no fragmentation along filaments due to the small-scale power spectrum cutoff, resulting in a more continuous distribution of dark matter. Furthermore, these filaments persist for longer periods in $\psi$DM, as the gravitational attraction from the early formation of the first halos in CDM tends to disrupt them earlier. The prolonged lifetimes of these filaments in wave dark matter favor the formation of a greater number of stars within them, leading to significant differences in the location and extent of stellar profiles in CDM and $\psi$DM galaxies. This also causes baryonic objects to appear more diffuse or smoothed compared to CDM, suggesting their potential as excellent tracers of dark matter in $\psi$DM \citep{Mocz:2019, Mocz:2020}.

 In dSph galaxies, stellar motions are predominantly influenced by the gravitational potential of dark matter ( they are much more dominated than giant galaxies and that's the main reason why are the ideal guinea pigs for this study), leading to anticipated observable differences in stars based on the nature of dark matter. The formation and early evolution of galaxies and filaments are also sensitive to the nature of dark matter, particularly influenced by the suppressed high-k power spectrum in Warm dark matter (WDM)/$\psi$DM models. The wave behavior and quantum pressure in $\psi$DM result in distinct virialized DM halo structures with prominent soliton cores, contrasting with the smooth cores seen in CDM. While CDM predicts cuspy profiles \cite{Navarro:1996, Lovell:2014}, simulations of merging DM haloes in $\psi$DM clearly reveal the formation of soliton structures at the de Broglie wavelength scale \cite{Schive:2014, Mocz:2017, Schwabe:2016}, standing out in terms of core density above the surrounding DM halo. This differs from the early fragmentation of filaments and cuspy haloes in CDM \cite{Mocz:2017, Mocz:2019, Mocz:2020}. Studies by \cite{Mocz:2019, Mocz:2020} demonstrate that baryonic feedback has limited impact on halos within the $10^{9}~M_{\odot}$ to $10^{10}~M_{\odot}$ range for redshifts $z>6$, failing to soften the cuspy profiles of CDM/WDM to produce cores. Consequently, neither CDM nor WDM can account for the claimed core origins from dynamical studies of dSph galaxies \citep{Read:2005, Amorisco:2013}.

This concept has been partially explored in prior studies \cite{Mocz:2019, Mocz:2020, Mocz:2017}, where simulations within the $\psi$DM framework have previously revealed the birth of stars along dense dark matter filaments, effectively tracing dark matter. This phenomenon has been emphasized as a distinctive ``smoking gun" signature of $\psi$DM \cite{Mocz:2019}. Additionally, it is worth noting that our recent work \cite{Pozo:2023} has further reinforced the notion that the anticipated stellar profiles from simulations for three major DM models—CDM, WDM, and $\psi$DM—exhibit slight but meaningful distinctions, directly linked to the unique characteristics of each individual DM model. Additionally, the analysis has been extended to the relative distribution of $\psi$DM and baryons in virialized galaxies. The conclusion is that, even though the baryonic profiles and dark matter profiles still differ, only $\psi$DM exhibits cases where a similar distribution can be found.

The presence of prominent cores is clearly observed in the star count profiles of both Ultra Faint Dwarf and dwarf Spheroidal galaxies, as illustrated in Figure \ref{figmeans}, when averaged within each class of dwarf. There is an evident difference in scale between UFD and dSph galaxies, with a radius factor of approximately 10. These cores are individually discernible in deep images, as presented in most cases (refer to the supplement for individual fits), and they are deemed "prominent" due to the core density rising well above the surrounding "halo" by a factor of 30 in density for both classes of dwarfs, as seen in Figure \ref{figmeans}. The stellar cores exhibit similarity to the commonly adopted Plummer profile (depicted by the red curve in Figure \ref{figmeans}), but they more accurately align with the soliton form of $\psi$DM in the ground state. This alignment is noteworthy despite the inherent parameter-free nature of the soliton profile, with the boson mass being the sole free parameter for $\psi$DM, determining the soliton radius. Furthermore, extended halos around these solitons are observed as a general feature, extending to kiloparsec scales, consistent with recent discoveries of halos around two well-studied dwarfs \cite{Chiti:2021,Collins:2021}. Such extended halos are intrinsic to $\psi$DM, composed of excited states above the ground state soliton, as indicated by the NFW form predicted by $\psi$DM simulations \cite{Schive:2014}. This reflects the inherently wave nature of $\psi$DM. The averaged core-halo structure of all dSph and UFD dwarfs is presented in Figure \ref{figmeans}, revealing a remarkably tight agreement. All individual profiles are showcased in the Supplement, highlighting the generality of this core-halo behavior across all well-studied dwarfs. This encompasses a typically sharp density transition between the core and the halo, as depicted in Figure \ref{figmeans} and observed in most individual dwarf profiles, indicated by vertical orange bands.

The remarkable agreement with the core-halo profile of $\psi$DM prompts the plotting of the core density versus radius relation for the soliton radius, individually measured for all dwarfs as shown in Figure \ref{figrcl}, where the velocity dispersion measured within that radius is also utilized. In Figure \ref{figrcl}, right panel, two parallel relations become more apparent for the Ultra Faint Galaxy and dwarf Spheroidal dwarfs, respectively, in terms of $\rho_c \propto \sigma^2 / R_c^2$. For $\psi$DM, a steep slope of $\rho_{sol} \propto R_{sol}^{-4}$ is predicted due to the volume dependence $R_{sol}^{-3}$ and the inherent inverse scaling of the soliton radius with soliton mass, $M_{sol} \propto 1/R_{sol}$, derived from the time-independent soliton solution of the Schrödinger-Poisson coupled equation (verified by $\psi$DM simulations \cite{Schive:2014,Schive:20142}). Consequently, $\sigma^2/ R_h^2 = \beta^2 (\hbar/m_b)^2/R_h^4$, where $\beta$ is an order unity dimensionless scaling provided by the Uncertainty Principle ($\hbar = m_b \sigma R_h$), resulting in a slope of $d\log\rho_{sol}/d\log R_{sol}= -4$ for $\psi$DM. This predicted slope aligns clearly with the data in Figure \ref{figrcl} (right panel) for both the Ultra Faint Dwarf and dwarf Spheroidal galaxies. This natural alignment explains the otherwise puzzling trend where large cores within each class of dwarf have lower velocity dispersions. Notably, the "feeble giant" dwarf, Crater II, adheres to this relation, as do estimates of the core radius for the Milky Way and the ultra-diffuse low mass galaxy DF44, measuring 50 pc and 120 pc, respectively \cite{deMartino:2018,Pozo:2021,vanDokkum:2019}. Importantly, this core density-radius relation remains unaffected by tidal stripping, inferred to have significantly affected Crater II and Antlia II with small pericenter orbits about the Milky Way. The stability of the soliton necessitates it always follows the inverse mass-radius relation set by the boson mass, allowing a stripped galaxy to move down the core density-radius relation without departing from it until the soliton is catastrophically destroyed by tidal forces \cite{Veltmaat:2018}.

Finally, we test directly for the role of the Uncertainty Principle to see if the inverse scaling is present between $\sigma_{c}$ and $r_{c}$ from the 
commutability of momentum and position required for the soliton. In Figure \ref{figrcd} we see this inverse relationship is indeed supported by both the UFD and dSph dwarfs, in parallel, which is quite the opposite of the positive correlation predicted for CDM \cite{Walker:2009} where more massive dwarfs are larger, as indicated by the red curve. This agreement means we can roughly estimate the boson mass for both the UFD and dSPh classes from the normalization between the core momentum $m_b \sigma_c$ and the width of the soliton standing wave so, $m_b=\hbar/2R_c \sigma_c$, fitted in Figure \ref{figrcl}, obtaining $m_b= 2.27^{+1.79}_{-0.65}\times 10^{-21}$~eV for the UFD's and $ m_b= 1.85^{+0.66}_{-0.58}\times 10^{-22}$~eV for the dSph's, differing by an order of magnitude. This simple estimate using the Uncertainty Principle may be compared to individual Jeans analysis of dSph dwarfs\cite{Chen:2017}, where a similar range of boson mass and core radius is derived dynamically for several dSph with high quality profiles, in the range $0.9-2.8\times 10^{-22}$eV. This estimate assumes that the stars and DM share the same spatial distribution, a reasonable assumption for stellar orbits that have relaxed over time \cite{Mocz:2017, Mocz:2019, Mocz:2020,Pozo:2023,DellaMonica:2023}, so the 3D velocity dispersion associated with the soliton wave function, where only the radial mode of kinetic energy $<KE>_r$ is present, means in 1D we have $r_c \sigma = 0.5 (\hbar/m)$, as adopted in our estimate. More precise absolute boson masses may need to rely on simulations as it is now clear that stellar orbit scattering by soliton oscillation modes affects the evolution of stellar orbits within the soliton \cite{DellaMonica:2023}. We emphasise that irrespective of absolute values, Figure \ref{figrcd} indicates there is an order of magnitude difference in boson mass between UFD and dSph dwarfs. Furthermore, this conclusion is supported independently by the dwarfs associated with the Milky Way and with Andromeda, prefixed by ``And" in Figures \ref{figrcl} \& \ref{figrcd}, for which we find indistinguishable core density relations and bosons masses, as listed in Table \ref{tabla:collage}, thus reinforcing the generality of our two boson solution for local dwarf galaxies. 

\begin{figure}[htp]
%\begin{varwidth}{\linewidth}
	\centering
    \includegraphics[scale=0.5]{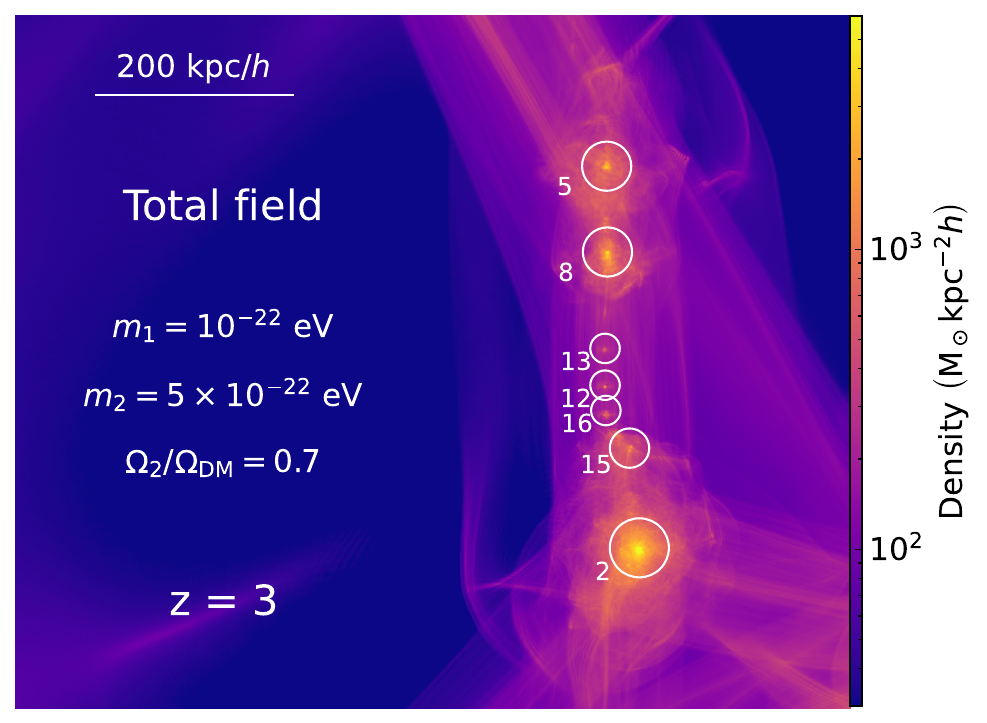} \\
    \includegraphics[scale=0.6]{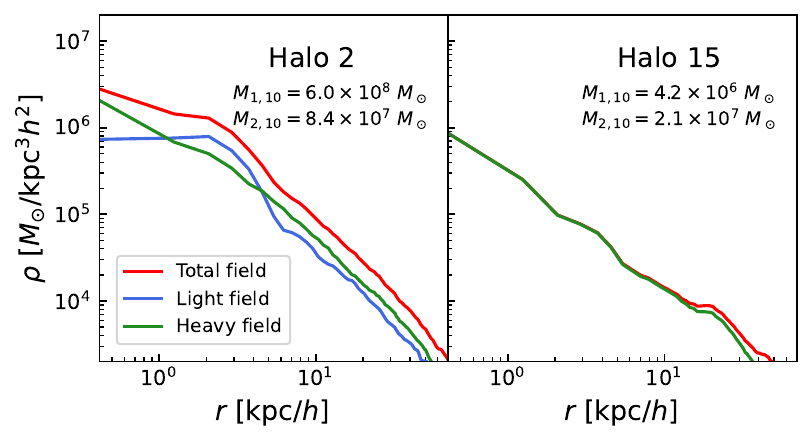}
    \caption{{\bf Representative filament from a 2-boson simulation}. Low-mass galaxies mainly form within the filaments and are dominated by the heavier boson (bottom left profile) as the relatively large de Broglie scale of the light boson prevents light boson fragmentation along filaments. The more massive halos are observed to form at the filament nodes with a mix of bosons (bottom right profile), reflecting the equal proportion of light and heavy bosons chosen for this simulation. The boson mass ratio $m_1/m_2=5$ here and $M_{10}$ denote the mass enclosed within 10 kpc of each halo.}\label{figsimu}
% \end{varwidth}  
\end{figure} 

Indeed, the assertion of two distinct populations of galaxies resulting from two different boson masses can only be accurate if some aspect of the galaxy formation process leads to the spatial separation of the two types of bosons. The initial simulations and studies aimed at addressing this question are relatively recent and have not definitively clarified it. For instance, \cite{Huang:2023} conducted simulations where they analyzed the evolution of galactic halos under gravity with two different populations of bosons, major and minor. They explored scenarios with a 75$\%$ major and 25$\%$ minor composition and vice versa. The results indicated that, on large scales, the spatial distributions of filaments and massive haloes were very similar between the two-component and single-component models, suggesting a comparable cosmological evolution that could support the coexistence of UFD and classical dwarfs. Additionally, they demonstrated how different proportions of major and minor components would affect the resulting core-halo structure. In cases where the major component dominated, a halo resulting from both bosons simultaneously would be possible. Both components contributed similarly to the soliton total density profile, resulting in a significantly lower soliton peak density than the single-component counterpart. This, along with the presence of an extended minor-component soliton, led to a much smoother soliton-to-halo transition. However, in scenarios where the minor component constituted 75$\%$  of the total population, they explicitly stated that the minor-component soliton could not form once the major-component soliton was stabilized \cite{Huang:2023}. These findings underscore that the evolution of different structures in a $\psi$DM context is feasible even with two distinct populations of bosons, but the spatial distribution and proportion of these populations are crucial considerations. Nevertheless, it is important to mention that The referenced paper by \cite{Huang:2023} shows that if there are two $\psi$DM components they will have different spatial distributions within a virialized halo but does not suggest that one can have different galaxies composed of different types of dark matter, which is a requirement for the objectives of this work. Unfortunately, no previous simulations have addressed this issue, making it a challenge that must be tackled in our future research.

On the other hand, there are many tensions between $\psi$DM predictions and observational data that can potentially be alleviated by a two-field model \cite{Nhan:2023}. For instance, the presence of multiple boson populations helps to smooth out the density fluctuations in $\psi$DM  haloes, thereby relaxing the constraints \cite{Gosenca:2023}. Additionally, a sufficiently large density fraction of the heavy field is expected to address the Lyman-$\alpha$ constraint \cite{Hui:2017, Nhan:2023}

Physically, the dominance of the heavy boson species occurs in the filament regions where the de Broglie scale of the light bosons is too large to allow fragmentation. Consequently, only low-mass halos are able to form, comprising the smaller de Broglie scale of the heavier boson (see Figure \ref{figsimu}). We also anticipate that the heavier galaxies formed at the nodes of filament intersection comprise approximately the initial mixture of heavy and light bosons, as depicted in Figure \ref{figsimu}. Simulations are underway to explore a range of boson mass ratios and relative initial boson densities (Luu, Mocz, Vogelsberger, Broadhurst, Pozo, Henry, Smoot, Emami $\&$ Hernquist 2024, in preparation) for a quantitative comparison with the observationally distinct UFD and dSph dwarf classes. This numerical investigation poses greater challenges than those explored to date, as it must encompass the factor of 20 in boson mass ratio preferred by our analysis in this paper.

Our two boson solution for dwarf galaxies may point to the ``Axiverse" scenario generic to string Theory \citep{Arvanitaki:2010}, where in general terms a wide spectrum of axion-like scalar fields is predicted with a discrete mass spectrum spanning many decades in mass, with approximately one axion per decade. In this context, we may infer that the proportion of the Universal DM in a higher mass boson may be approximately $\sim$3\% compared to the lighter boson, given the factor 10 mass difference we find, as the higher mass axion enters the horizon earlier and is redshifted to lower density ahead of lower mass axion \citep{Luu:2020}. Our two boson conclusion improves the viability of this String Theory solution for DM raised in relation to the existence of UFD galaxies \citep{Mohammadtaher:2020} and may relieve tension with $\psi$DM based on excessive variance of the of the Ly-forest \citep{Armengaud:2017}, though gas outflows and early AGN heating must also be expected to enhance the forest variance above ideal DM simulation based predictions at some level, as too may an initially ``extreme" angle scalar field for $\psi$DM \citep{Hsu:2021}. We can also now anticipate constraints on this two boson solution from JWST, where early galaxy formation related to the subdominant, heavier boson will be governed by the dominant density field of the lighter boson and hence strongly biassed, favouring the formation of UFD dwarfs in groups and clusters. Alternatively, JWST may reveal that dwarf galaxies are physically continuous at early times, as expected for scale free CDM, implying subsequent evolutionary processes are responsible for the physical distinction between UFD and dSph dwarfs seen today.

\begin{table}[h!]
	\centering

\begin{tabular}{|c|c|c|c|c|}

\hline

 \multirow{2}{*}{Combinations}& $r_c$  & $r_{t}$& \multirow{2}{*}{$N_{\rm gal}$} &   $m_{\psi}$  \\
& (kpc)  & (kpc) & & $(10^{-22}$~eV)\\
\hline

${\rm dSph}_{\rm both}$  &$0.21^{+0.003}_{-0.003}$ &$0.71^{+0.021}_{-0.021}$ &23 &$1.85^{+0.66}_{-0.58}$\\
\hline

${\rm UFD}_{\rm both}$  &$0.033^{+0.002}_{-0.002}$ &$0.11^{+0.006}_{-0.006}$ &21 &$23.21^{+17.91}_{-7.23}$\\
\hline

${\rm dSph}_{\rm Milky~Way}$  & $0.22^{+0.003}_{-0.003}$&$0.75^{+0.022}_{-0.023}$ &13 &$1.85^{+0.66}_{-0.58}$\\
\hline

${\rm dSph}_{\rm Andromeda}$  &$0.26^{+0.007}_{-0.006}$ &$0.82^{+0.032}_{-0.028}$ &10 &$1.86^{+0.45}_{-0.53}$\\
\hline

${\rm UFD}_{\rm Milky~Way}$  &$0.032^{+0.002}_{-0.002}$ & $0.093^{+0.008}_{-0.007}$&12 &$31.36^{+12.39}_{-11.57}$\\
\hline

${\rm UFD}_{\rm Andromeda}$  &$0.042^{+0.002}_{-0.002}$ &$0.14^{+0.008}_{-0.008}$ &9 &$17.83^{+4.31}_{-6.73}$\\
\hline

\hline

\hline

\end{tabular}
\caption{Profile parameters for dwarfs associated with the Milky Way and Andromeda. Column 1: Dwarf Class, Column 2: Core radius  $r_c$,  Column 3: Core-Halo transition radius $r_t$, Column 4: Number of galaxies $N_{\rm gal}$, Column 5: Boson mass $m_{\psi}$.}

\label{tabla:collage}
\end{table}

\bibliography{apssamp}% Produces the bibliography via BibTeX.

\begin{acknowledgments}
We thank Justin Schive for useful comments
and for providing the simulation profiles.
T. Broadhurst and H. N. Luu are supported by the Collaborative Research Fund under Grant No. C6017-20G which is issued by Research Grants Council of Hong Kong S.A.R.
\break
\end{acknowledgments}

\newpage
\onecolumngrid
%\section*{Appendix}
\appendix
\hfill \break

\begin{center}
    
\textbf{\huge Methods:}

\end{center}

\section{The Wave Dark Matter Halo}\label{wavehalo}

The light bosons paradigm was first introduced by \cite{Widrow:1993}, \cite{Sahni:2000} and  \cite{Hu:2000}, and subsequently re-considered with the first simulations \cite{Marsh:2014, Schive:2014, Bozek:2015,Hui:2017,Veltmaat:2018,Robles:2019,Niemeyer:2020} and in relation to the puzzling properties of dwarf spheroidal galaxies. In the simplest version, without self-interaction, the boson mass, $m_b$, is the only free parameter, with a fiducial value of $10^{-22}$ eV adopted to match the approximate Kpc scale commonly reported for dark matter dominated dwarf galaxy cores. \hfill \break

The first simulations in this context have revealed a surprisingly rich wave-like structure with a solitonic standing wave core, surrounded by a halo of interference that is fully modulated on the de Broglie scale \cite{Schive:2014}. The solitonic core corresponds to the ground-state solution of the coupled Schr\"oedinger-Poisson equations, with a cored density profile well-approximated by \cite{Schive:2014, Schive:20142}
\begin{equation}\label{eq:sol_density}
\rho_c(r) \sim \frac{1.9~a^{-1}(m_\psi/10^{-23}~{\rm eV})^{-2}(r_{c}/{\rm kpc})^{-4}}{[1+9.1\times10^{-2}(r/r_{c})^2]^8}~M_\odot {\rm pc}^{-3}\,,
\end{equation}
Here $m_\psi$ is the boson mass, and $r_{c}$ is the solitonic core radius, which simulations show scales as halo mass \cite{Schive:20142} in the following way: 
\begin{equation}\label{eq:rc_c}
    r_{c} \propto m_\psi^{-1}M_{halo}^{-1/3}\,.
\end{equation}

\begin{equation}\label{eq:sol_radius}
r_c=1.6\biggl(\frac{10^{-22}}{m_\psi}  eV \biggr)a^{1/2}
\biggl(\frac{\zeta(z)}{\zeta(0)}\biggr)^{-1/6}
\biggl(\frac{M_h}{10^9M_\odot}\biggr)^{-1/3}kpc
\end{equation}

Core masses of constant density scale as $\rho_c \propto (\sigma/r_c)^2$ and
in the context of $\psi$DM there is also an inverse relationship between soliton core 
mass and soliton radius relation required by the non-linear solution to the Schrodinger-Poisson equation\cite{Schive:2014} so the soliton's density scales more steeply than the volume with radius, i.e. $\rho_c \propto r_c^{-4}$. The radius of the soliton is given approximately by the de Broglie wavelength $\lambda_{B} =\frac{h}{p}$, following from the Uncertain principle $\Delta{x}\Delta{p} \geq\frac{\hbar}{2}$, Where $\Delta{x}$, the position dispersion given by the soliton width, 2$\times r_{c}$, and the dispersion in momentum $\Delta{p}$, given approximately by $m_b \sigma$, the product of the boson mass and the velocity dispersion of stars as tracer particles of the dominant DM potential. This allow us to determine the boson mass that corresponds to the de Broglie wavelength, $m_{\psi}\simeq {\hbar}$/$4 r_{c} \sigma_{los}$.\hfill \break

The simulations also show the soliton core is surrounded by an extended halo of density fluctuations on the de Broglie scale that arise by self interference of the wave function \cite{Schive:2014} and is ``hydrogenic" in form \cite{Hui:2017,Vicens:2018}. These cellular fluctuations are large, with full density modulation on the de Broglie scale \cite{Schive:2014} that modulate the amplitude of the Compton frequency oscillation of the coherent bosonic field, allowing a direct detection via pulsar timing \cite{deMartino:2017,deMartino:2018}.\hfill \break

This extended halo region, when azimuthally averaged, is found to follow the Navarro-Frank-White (NFW) density profile \cite{Navarro:1996, Woo:2009,Schive:2014, Schive:20142}
%Nevertheless, this feature becomes important for phenomena happening of timescale less than few months allowing, for example, the detection of ultralight bosons with current and/or forthcoming pulsar timing array experiments \cite{idm2017,idm2018}. 
%The fitting formula for the density profile of the solitonic core in a $\psi$DM halo  is obtained from cosmological simulation 
so that the full radial profile may be approximated as:\hfill \break
\begin{equation}\label{eq:dm_density}
\rho_{DM}(r) =
\begin{cases} 
\rho_c(x)  & \text{if \quad}  r< Xr_{c}, \\\\
\frac{\rho_0}{\frac{r}{r_s}\bigl(1+\frac{r}{r_s}\bigr)^2} & \text{otherwise,}
\end{cases}
\end{equation} 
where $\rho_0$ is chosen such that the inner solitonic profile 
matches the outer NFW-like profile at approximately $\simeq Xr_{c}$, and $r_s$ is the scale radius.\hfill \break

In this context, we can now predict the corresponding velocity dispersion profile by solving the spherically symmetric Jeans equation:\hfill \break
\begin{equation}\label{eq:sol_Jeans}
\frac{d(\rho_*(r)\sigma_r^2(r))}{dr} = -\rho_*(r)\frac{d\Phi_{DM}(r)}{dr}-2\beta\frac{\rho_*(r)\sigma_r^2(r)}{r},
\end{equation}
where $\rho_*(r)$ is the stellar density distribution is the stellar density distribution defined by the solitonic wave dark matter profile:

\begin{equation}\label{eq:stellar_density}
\rho_{*}(r) =
\begin{cases} 
\rho_{1*}(r)  & \text{if \quad}  r< r_t, \\
\frac{\rho_{02*}}{\frac{r}{r_{s*}}\bigl(1+\frac{r}{r_{s*}}\bigr)^2} & \text{otherwise,}
\end{cases}
\end{equation}
where
\begin{equation}\label{eq:stellar_1}
\rho_{1*}(r) = \frac{\rho_{0*}}{[1+9.1\times10^{-2}(r/r_c)^2]^8}~N_* {\rm kpc}^{-3}
\end{equation}

Here, $r_{s*}$ is the 3D scale radius of the stellar halo corresponding to $\rho_{0*}$ the central stellar density, $\rho_{02*}$ is the normalization of $\rho_{0*}$ at the transition radius and the transition radius, $r_t$, is the point where the soliton structure ends and the halo begins at the juncture of the core and halo profiles. $\beta$ is the anisotropy parameter, defined as (see \cite{Binney:2008}, Equation (4.61))
\begin{equation}\label{eq:sol_beta}
\beta = 1 - \frac{\sigma_t^2 }{\sigma_r^2}.
\end{equation}
%\beta = 1 - \frac{\sigma_\phi^2 + \sigma_\theta^2}{2\sigma_r^2}\,.

%Since the galaxy may be decomposed as a $\psi$DM halo plus a sub-dominant stellar component, the dynamics is fully determined by the potential well of the DM halo.
Thus, the gravitational potential is given by:
\begin{equation}\label{eq:potbulge1}
d\Phi_{DM}(r) = G \frac{M_{DM}(r)}{r^2} dr\,,
\end{equation}
with a boundary condition  $\Phi_{DM}(\infty)=0$, and the mass enclosed in a sphere of radius $r$ is computed as follows
\begin{equation}\label{eq:massbulge1}
M_{DM}(r) = 4\pi\int_0^r x^2 \rho_{DM}(x)dx\,.
\end{equation}

Finally, to directly compare our predicted dispersion velocity profile with the observations, we have to project the solution of the Jeans equation along the line of sight as follows:
\begin{equation}\label{eq:sol_projected}
\sigma^2_{los} (R) = \frac{2 }{\Sigma(R)} \int_{R}^{\infty} \biggl(1-\beta \frac{R^2}{r^2}\biggr) \frac{\sigma^2_r(r)\rho_*(r)}{(r^2-R^2)^{1/2}} r dr\,\,
\end{equation}
where 
\begin{equation}
\Sigma(R) =2\int_{R}^{\infty} \rho_*(r)(r^2-R^2)^{-1/2}rdr\,. 
\end{equation}

\newpage

\section{Data Analysis and Results}

Here, we present the stellar density profiles of a comprehensive sample of dSph and UFD dwarfs, which we compare with the generic $\psi$DM core-halo profile outlined in Section \ref{wavehalo} (Equation \ref{eq:stellar_density}). As evident in all the figures in Sections \ref{classical} and \ref{UFD}, these dwarfs indeed exhibit a distinct common form—a core-halo structure similar to what is predicted for $\psi$DM. The cores consistently adhere to the unique soliton form in cases with deep star counts. Additionally, the azimuthally averaged outer region at a larger radius is well-fitted by the NFW profile, as anticipated for $\psi$DM \cite{Schive:2014}. The core and halo regimes are easily distinguishable, as the core exhibits higher density compared to the halo. The transition radius is marked by an orange vertical line in the plots. These figures illustrate that this profile behavior is consistent for dwarfs orbiting both Andromeda and the Milky Way, regardless of whether they are classified as "ultra-faint" or "dwarf Spheroidal." The extension of these NFW-like stellar halos can be traced in some dwarfs to over 2 kpcs in radius. In contrast, the cores are typically 0.5 kpc for dSphs and an order of magnitude smaller on average for UFDs, at 0.05 kpc.\hfill\break

%Here we show all the stellar density profiles of a comprehensive sample of dSph and UFD dwarfs that we compare with the generic $\psi$DM core-halo profile of section \ref{wavehalo} (Equation \ref{eq:stellar_density}). As can be seen in all the figures of sections \ref{classical} and \ref{UFD}, these dwarfs do appear to have a distinctive common form, a similar core-halo structure predicted for $\psi$DM. The cores accurately conform to the unique soliton form in all cases where the star counts are deep, and also the azimuthally averaged outer at a larger radius is well fitted by the NFW profile as predicted for $\psi$DM\cite{Schive:2014}. These core and halo regimes are distinct because the core is prominent in density above the halo, with the orange vertical line in the plots marking the transition radius. The figures make clear that this profile behaviour is similar for dwarfs orbiting Andromeda's and the Milky Way, which are classified as either ``ultra-faint" or ``dwarf Spheroidal". The extension of these NFW-like stellar halos is traceable in some dwarfs to over 2 kpcs in radius, whereas the cores are typically 0.5kpc for the dSPh's and an order of magnitude smaller on average for the UFD's, 0.05 kpc.\hfill\break

%\begin{figure}[h]
%\begin{minipage}[h]{0.49\linewidth}
%\includegraphics[width=1\linewidth]{ConerMeanALL1.png}
%\end{minipage}
%\hfill
%\begin{minipage}[h]{0.49\linewidth}
%\includegraphics[width=1\linewidth]{ConerMeanALL2.png}
%\end{minipage}
%\caption{}
%\label{ris:image1}

%\end{figure}

%\vfill
%\noindent

\begin{figure*}[h!]
  
  \centering
  \includegraphics[width=1\textwidth,height=13cm]{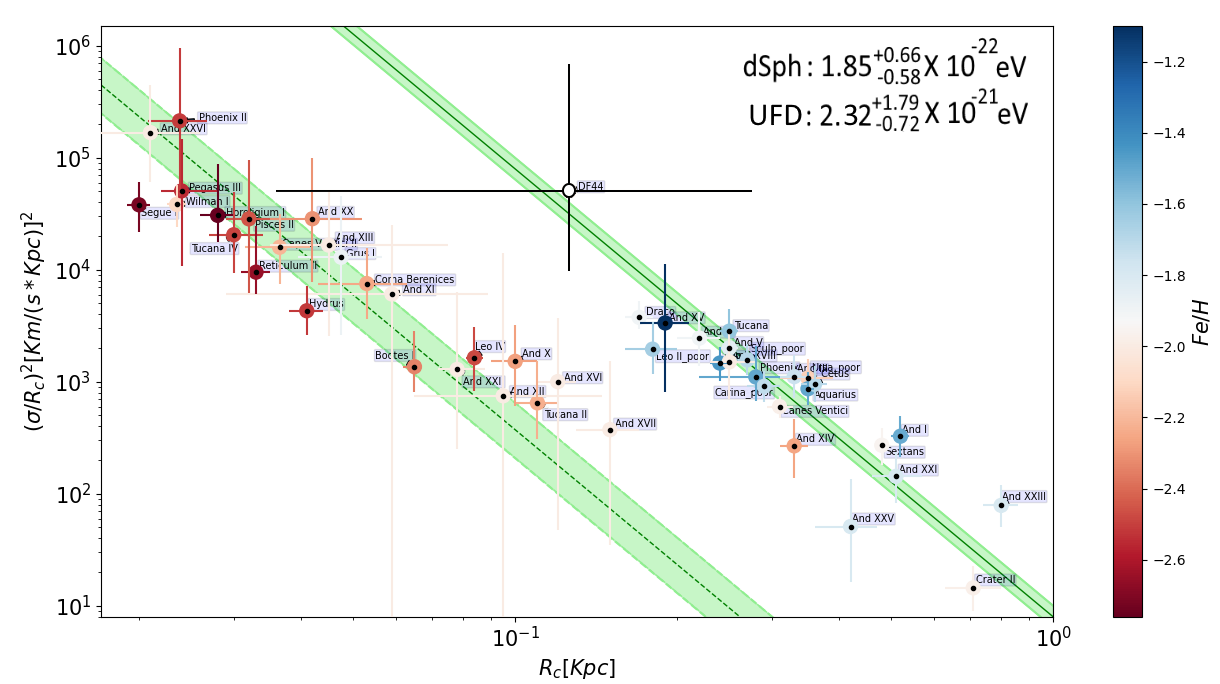}
  \caption{{\bf DM density vs. Core radius}. Expanding upon the left panel of Figure \ref{figrcl}, we now incorporate color to represent metallicity. It is noteworthy that the Ultra-faint galaxies consistently exhibit lower metallicity compared to the dSph class, thus reinforcing the empirical distinction between these two classes of dwarf galaxies, which is based on luminosity.}\label{figrcl2}
\end{figure*}

\begin{figure*}[h!]
  
  \centering
  \includegraphics[width=0.9\textwidth,height=10cm]{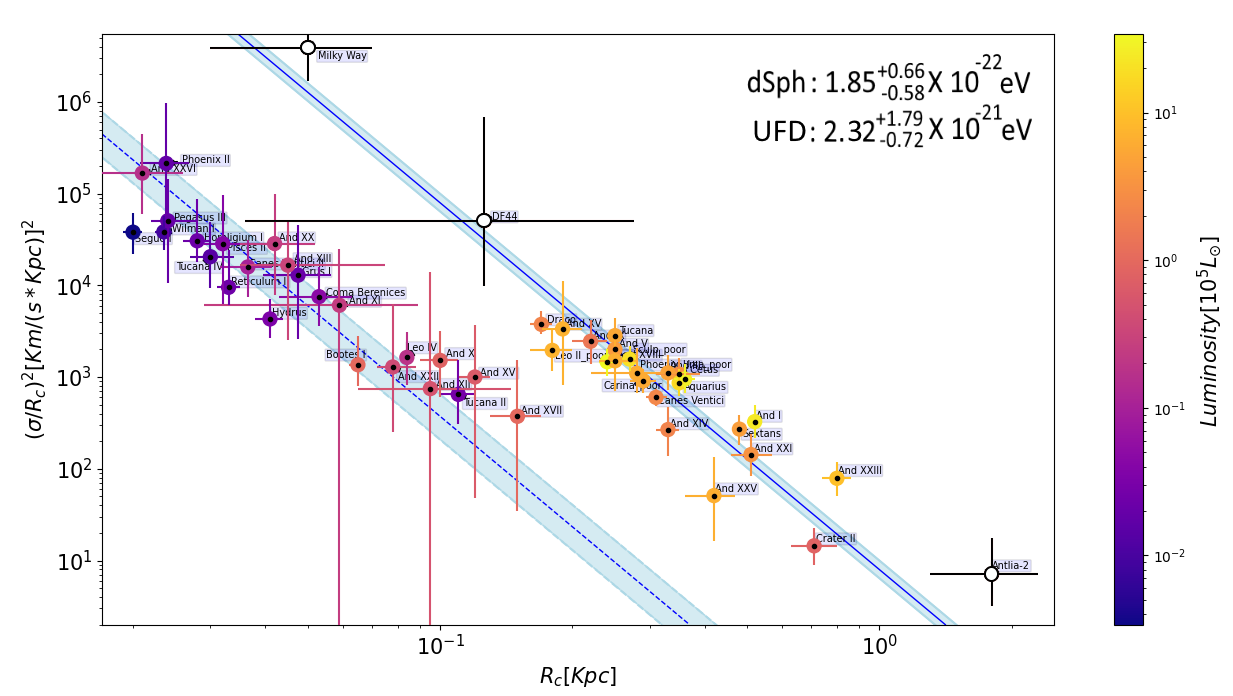}
  \caption{{\bf DM density vs Core radius}. Expansion of the right panel of Fig.\ref{figrcl}.}\label{figrc2}
\end{figure*}

\begin{figure*}[h!]
  
  \centering
  \includegraphics[width=0.9\textwidth,height=10cm]{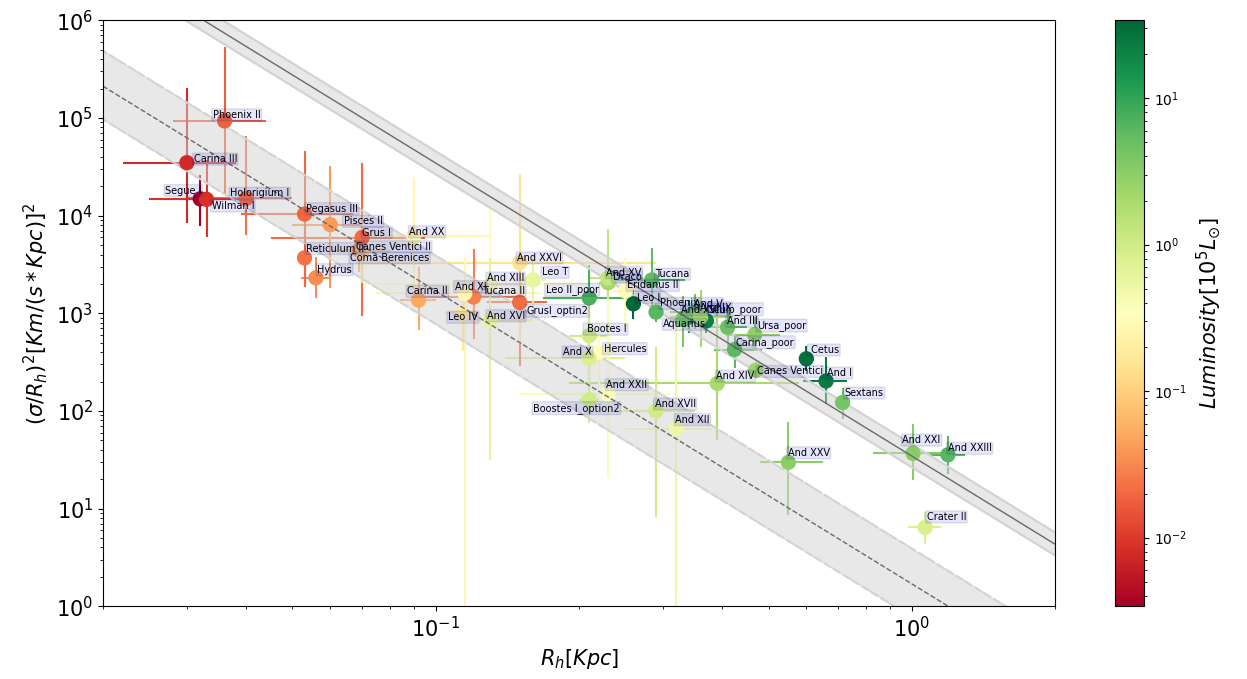}
  \caption{{\bf DM density vs Half-light radius}. Expansion of the left panel of Fig.\ref{figrcl}.}\label{figrh}
\end{figure*}

\begin{figure*}[h!]
  %\strut\newline
  \centering
  \includegraphics[width=0.8\textwidth,height=9cm]{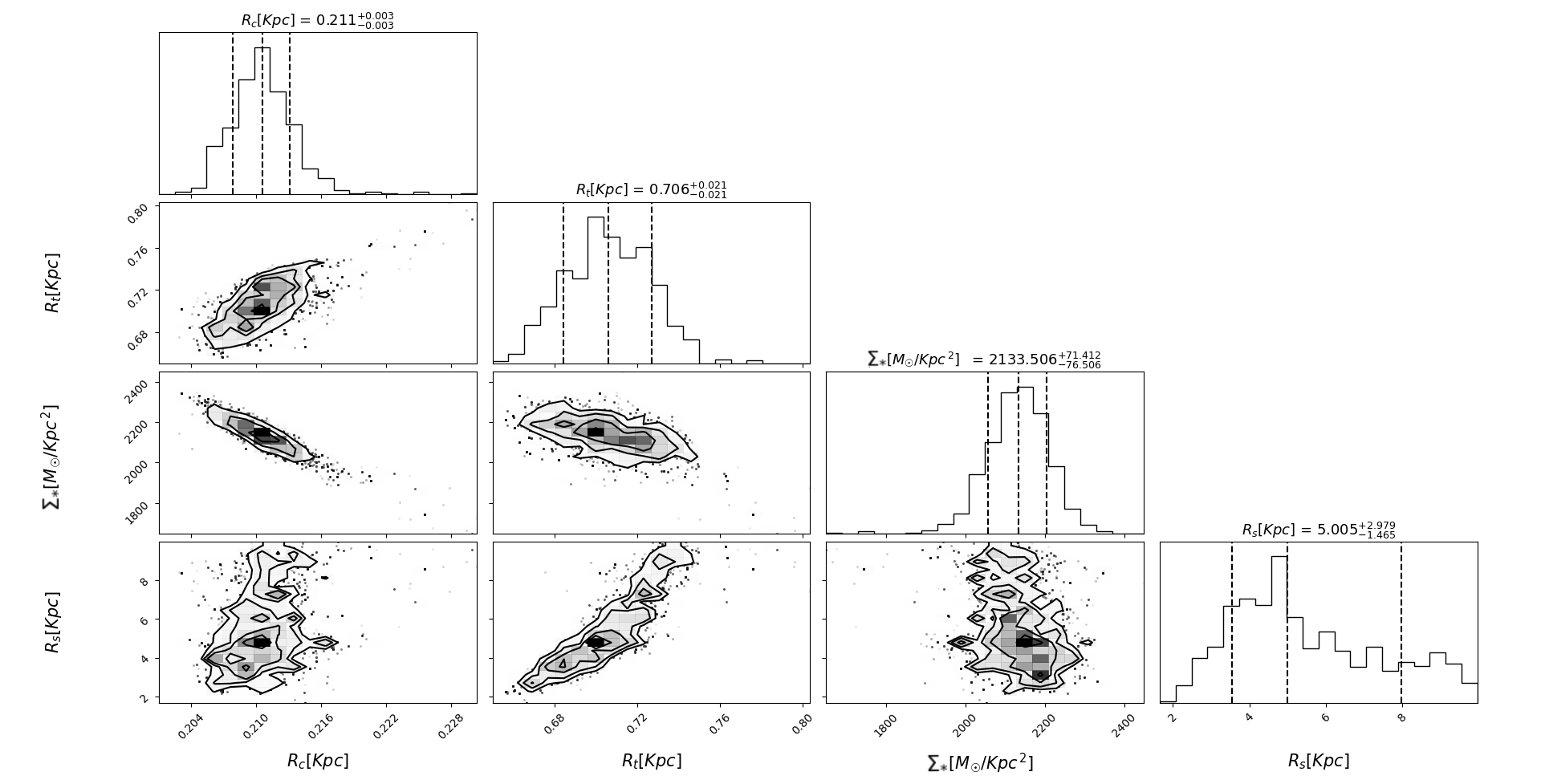}
  \caption{\textbf{All dSph}: Classical dwarfs mean profile( Figure \ref{figmeans} top panel): correlated distributions of the free parameters.  As can be seen the core radius and transition radius are well defined despite wide Gaussian priors, indicating a reliable result. The contours represent the 68\%, 95\%, and 99\% confidence levels. The best-fit parameter values are the medians (with errors), represented by the dashed black ones, and tabulated in Table \ref{tabla:collage}.}\label{corner1meanall}
\end{figure*}

\begin{figure*}[h!]
  
  \centering
  \includegraphics[width=0.8\textwidth,height=9cm]{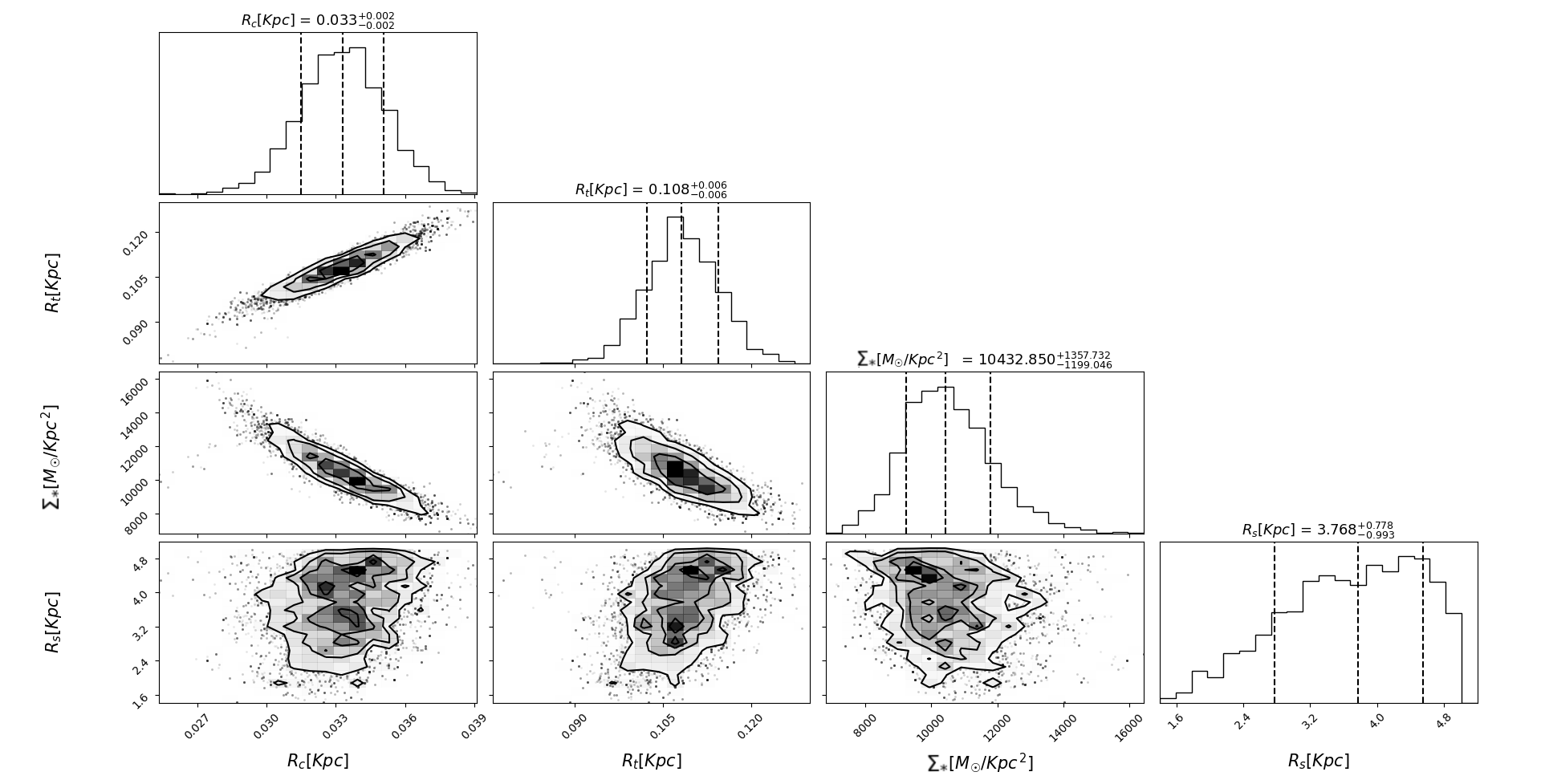}
  \caption{\textbf{All UFD }:Ultra faint dwarfs mean profiles( Figure \ref{figmeans} low panel): correlated distributions of the free parameters.  As can be seen the core radius and transition radius are well defined despite wide Gaussian priors, indicating a reliable result. The contours represent the 68\%, 95\%, and 99\% confidence levels. The best-fit parameter values are the medians(with errors), represented by the dashed black curve, and tabulated in Table \ref{tabla:collage}.}\label{corner2meanall}
\end{figure*}

\newpage

\subsection{ Classical Dwarf Galaxies}\label{classical}
%\title{Classical Dwarf Galaxies}\label{classical}

Analysis of the core-halo structure is conducted for all dSph stellar densities. Our predictions for the dSph class ($1.5 \times 10^{-22}$eV) in $\psi$DM are illustrated in green, representing the 2$\sigma$ range of the posterior distribution of profiles. We have included nearly all the dSphs within the Local Group, incorporating stellar profile data points beyond 1.5 kpc. It is crucial to emphasize that all these galaxies are consistent with the core-halo structure, both in the Milky Way and Andromeda, underscoring the universality of this structural pattern for dwarf galaxies.

\begin{figure*}[h!]
  
  \centering
  \includegraphics[width=0.8\textwidth,height=14cm]{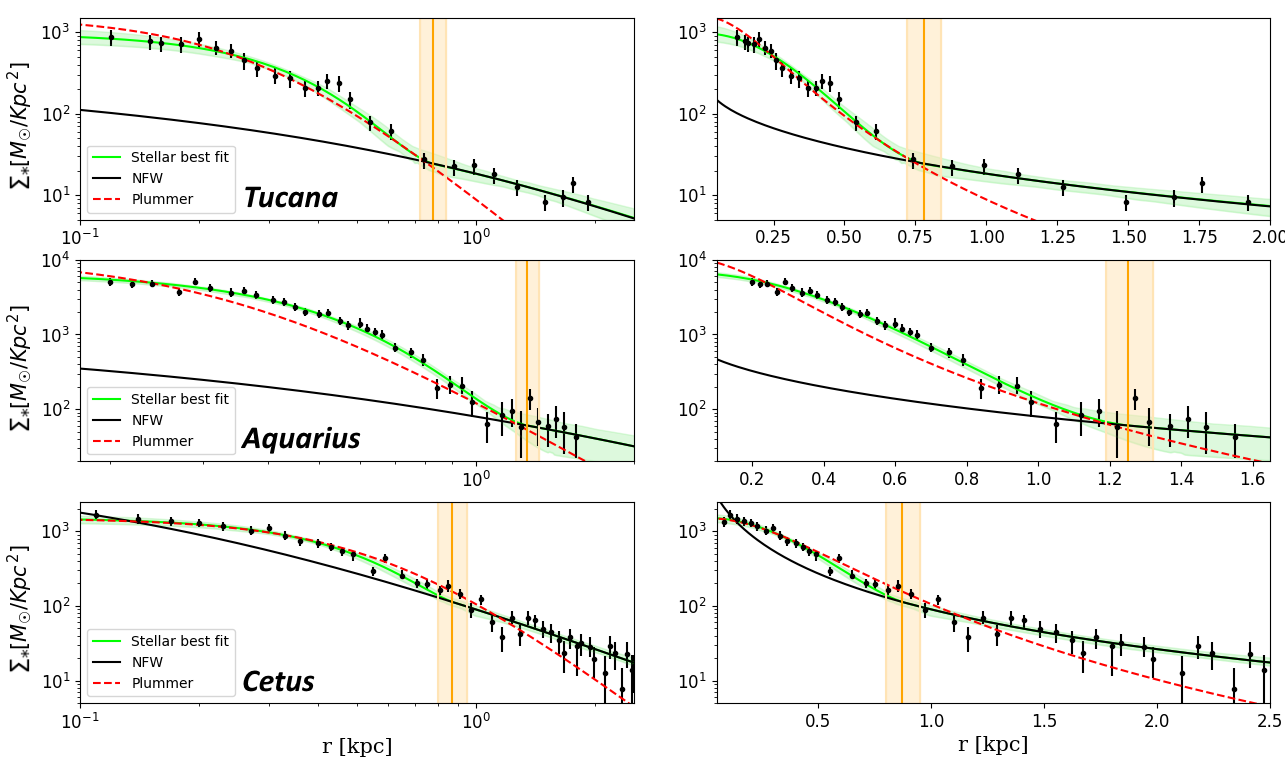}
  \caption{{\bf Dwarf Spheroidal Galaxies:} This figure presents star count profiles versus dwarf galaxy radius for well-studied dSph dwarf galaxies in the local group, as listed in Table \ref{tabla:1}. In most cases, an extended halo of stars is observable, stretching to approximately $\simeq 2$ kpc, most prominently visible on the linear scale of the left-hand panel. Prominent cores are also evident on a scale of less than 1 kpc in each dwarf. A standard Plummer profile (red dashed curve) roughly fits the core region but falls significantly short at larger radii. Our predictions for the dSph class ($\simeq10^{-22}$ eV) in $\psi$DM are depicted in green, where the distinctive soliton profile provides an excellent fit to the observed cores and the surrounding halo of excited states that azimuthally average to an approximately NFW-like profile beyond the soliton radius. The accuracy of the core fit to the soliton is best observed on a log scale in the right panels, while the left panel in linear scale shows the extent of the halo. This includes the characteristic density drop of about a factor of $\simeq 30$ predicted by $\psi$DM between the prominent core and tenuous halo at a radius of approximately 1 kpc, indicated by the vertical orange band. The best-fit MCMC profile parameters are tabulated in the supplement, and references to the data in this figure are as follows: Tucana \cite{Gregory:2019}, Cetus \cite{McConnachie:2006}, and Aquarius \cite{McConnachie:20062}.}\label{fig1}
\end{figure*}

\begin{figure*}[h!]
  
  \centering
  \includegraphics[width=0.8\textwidth,height=9.5cm]{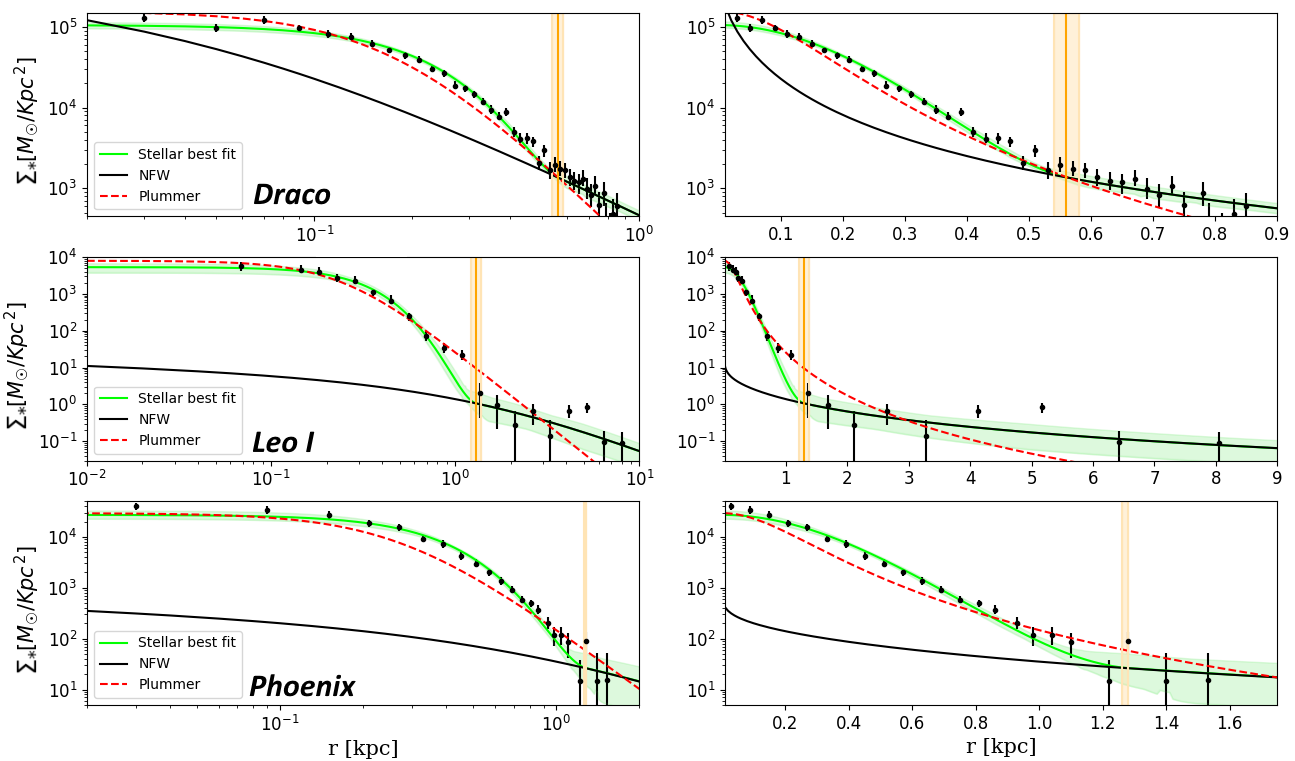}
  \caption{{\bf Dwarf Spheroidal Galaxies:} Similar to Figure \ref{fig1}, this figure includes three additional dSph galaxies. References to the data for these galaxies are as follows: Draco \cite{Wilkinson:2004}, Leo I \cite{Sohn:2007}, and Phoenix \cite{Battaglia:2012}.}\label{fig7}
\end{figure*}

\begin{figure*}[h!]
  
  \centering
  \includegraphics[width=0.8\textwidth,height=9.5cm]{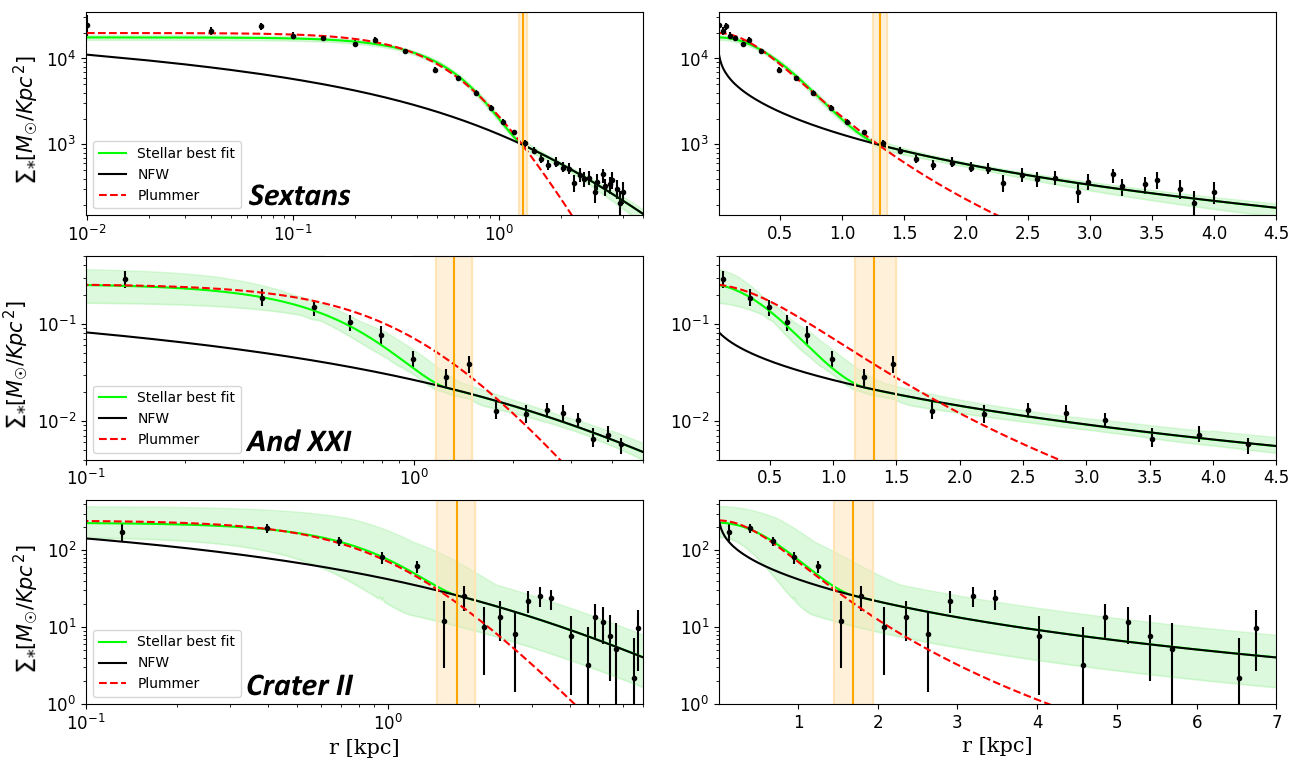}
  \caption{{\bf Dwarf Spheroidal Galaxies:}  
In line with Figure \ref{fig1}, this figure features three additional dSph galaxies. References for the data are as follows: Sextans \cite{Okamoto:2017}, Andromeda XXI \cite{Collins:2021}, and Crater II \cite{Torrealba:2016}. It is noteworthy to highlight that Andromeda XXI exhibits the same $\psi$DM core-halo structure as the dSph satellites of the Milky Way, further supporting the "universality" of this profile for dwarfs.}\label{fig8}
\end{figure*}

\begin{figure*}[h!]
  
  \centering
  \includegraphics[width=0.8\textwidth,height=10cm]{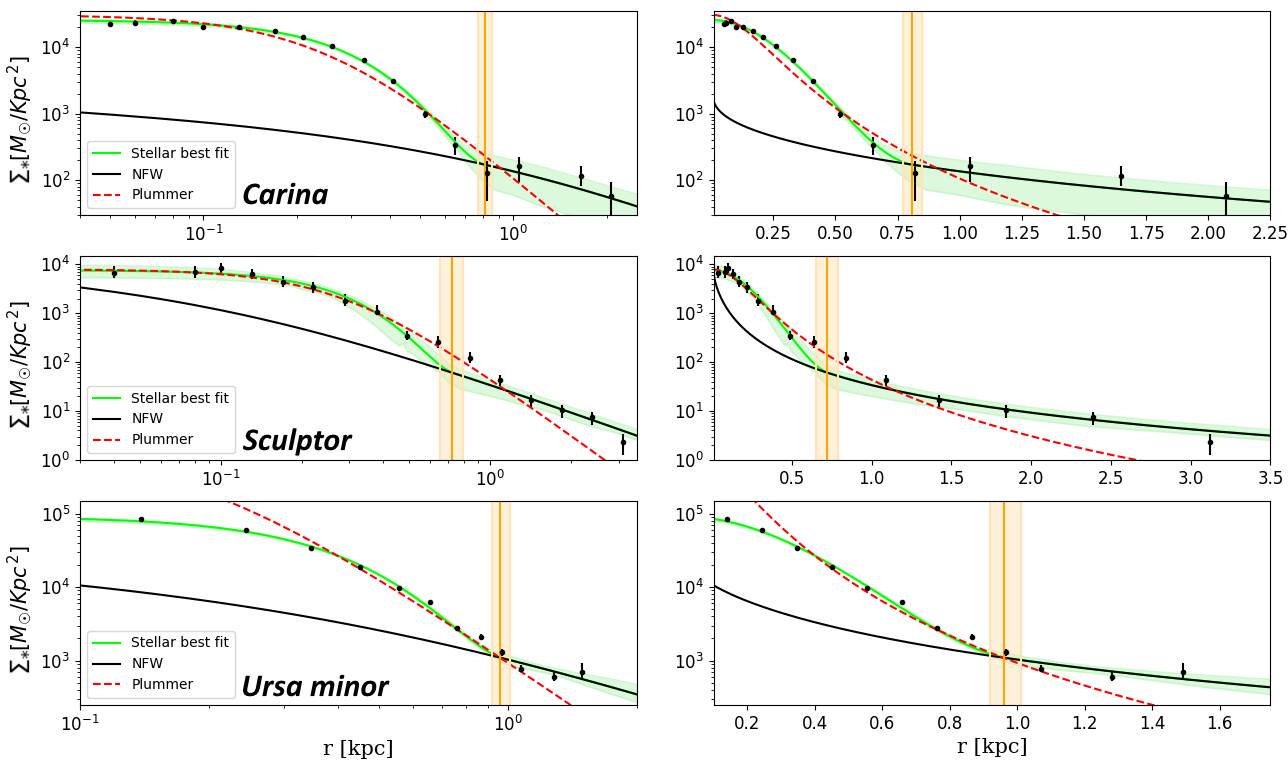}
  \caption{{\bf Dwarf Spheroidal Galaxies:}  As in Figure \ref{fig1}, this figure includes three additional galaxies. References for the data are as follows: Carina \cite{Frinchaboy:2012}, Sculptor \cite{Frinchaboy:2012}, and Ursa Minor \cite{Martinez-Delgado:2001}.}\label{fig9}
\end{figure*}

\begin{figure*}[h!]
  
  \centering
  \includegraphics[width=0.8\textwidth,height=10cm]{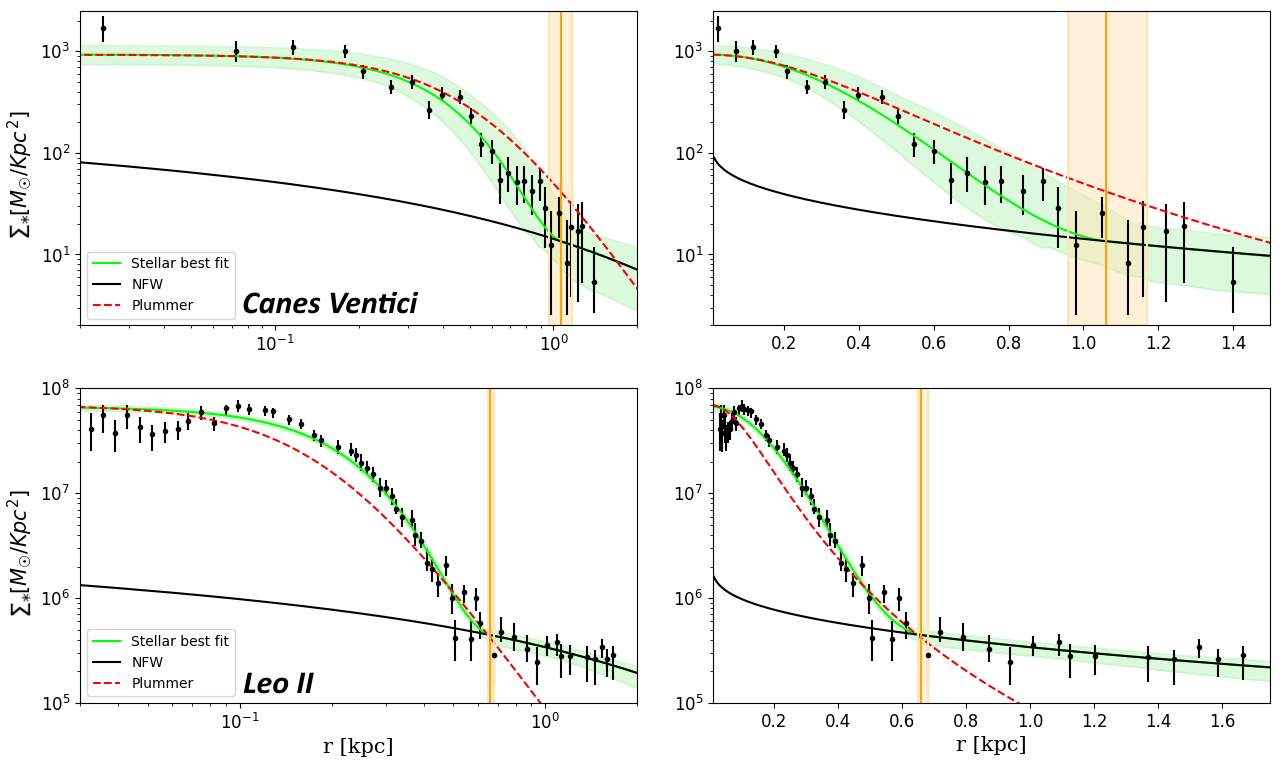}
  \caption{{\bf Dwarf Spheroidal Galaxies:}  
Similar to Figure \ref{fig1}, this figure showcases three additional galaxies. References for the data are Canes Venatici \cite{Zucker:2006} and Leo II \cite{Moskowitz:2020}.}\label{fig10}
\end{figure*}

\begin{figure*}[h!]
  
  \centering
  \includegraphics[width=0.8\textwidth,height=9.5cm]{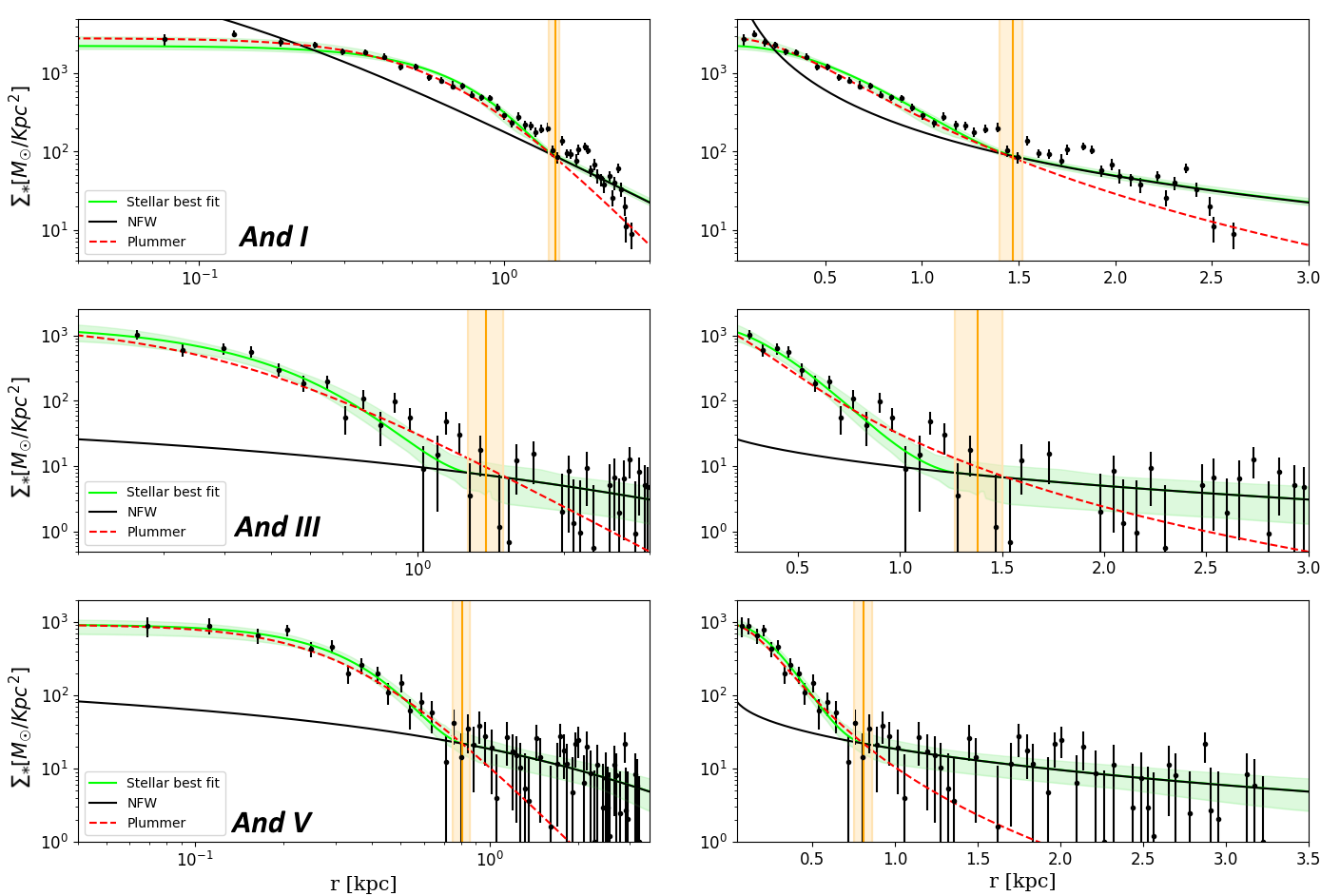}
  \caption{{\bf Dwarf Spheroidal Galaxies:}  Similar to Figure \ref{fig1}, this figure includes three additional galaxies. Notably, these Andromeda galaxies exhibit the same $\psi$DM core-halo structure as UFD galaxies in the Milky Way, underscoring the universality of the $\psi$DM profile for dwarfs. References for the data are as follows: Andromeda I \cite{Saremi:2020}, Andromeda III \cite{Martin:2016}, and Andromeda V \cite{Martin:2016}.}\label{fig11}
\end{figure*}

\begin{figure*}[h!]
  
  \centering
  \includegraphics[width=0.8\textwidth,height=9.5cm]{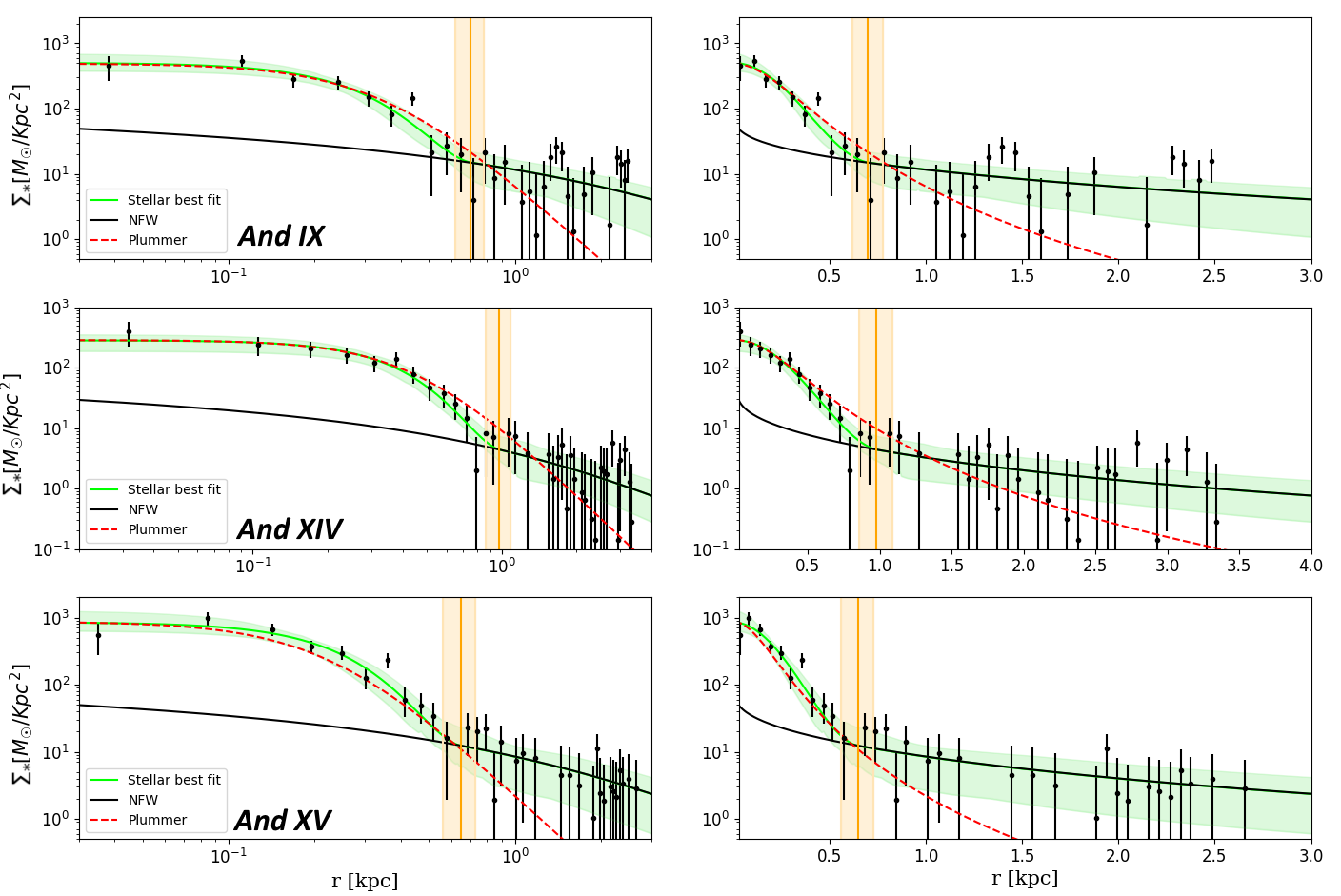}
  \caption{{\bf Dwarf Spheroidal Galaxies:}  
As in Figure \ref{fig1}, this figure includes three additional galaxies. References for the data are: Andromeda IX \cite{Martin:2016}, Andromeda XIV \cite{Martin:2016}, and Andromeda XV \cite{Martin:2016}.}\label{fig12}
\end{figure*}

\begin{figure*}[h!]
  
  \centering
  \includegraphics[width=0.7\textwidth,height=8cm]{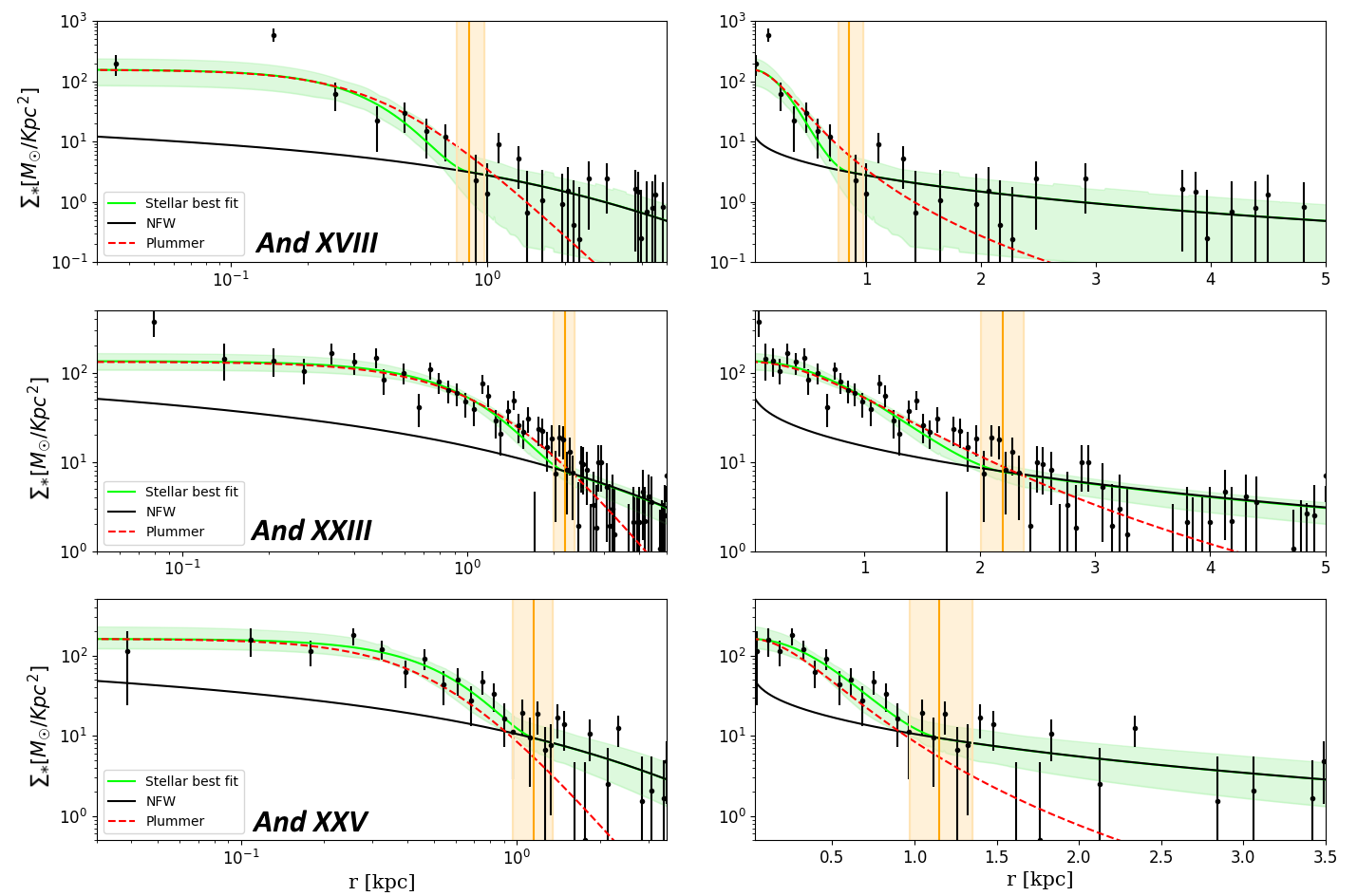}
  \caption{{\bf Dwarf Spheroidal Galaxies:}  
Similar to Figure \ref{fig1}, this figure includes three additional galaxies. References for the data are: Andromeda XVIII \cite{Martin:2016}, Andromeda XXIII \cite{Martin:2016}, and Andromeda XXV \cite{Martin:2016}.}\label{fig13}
\end{figure*}

\begin{table*}[h!]
	\centering

\begin{tabular}{|c|c|c|c|c|c|c|c|}

\hline

 Galaxy& $r_c$  & $r_{t}$& $r_{s*}$ &   $\sigma_{los,obs}$ &$r_{half,obs}$& $L_{obs}$& [Fe/H],obs \\
&(kpc)&  (kpc) &(kpc)  & (km/s)&(kpc)&$(10^{5}L_\odot)$&\\
\hline

Tucana  &$0.25^{+0.01}_{-0.01}$ &$0.78^{+0.06}_{-0.06}$   &$1.05^{+0.50}_{-0.57}$ &$13.3^{+2.7}_{-2.3}$\cite{Gregory:2019} &$0.284^{+0.05}_{-0.05}$\cite{Gregory:2019}&5.5\cite{Gregory:2019} &$\sim$-1.6\cite{Taibi:2020}\\
\hline

Cetus  & $0.36^{+0.02}_{-0.02}$ &$0.87^{+0.08}_{-0.07}$&$0.24^{+0.14}_{-0.06}$ & $11.1^{+1.6}_{-1.3}$\cite{Taibi:2018}&$0.6^{+0.01}_{-0.01}$\cite{Taibi:2018} &$28^{+8}_{-8}$\cite{Taibi:2018} &$\sim$-1.7\cite{Taibi:2018}\\
\hline

Aquarius  &$0.35^{+0.01}_{-0.01}$ & $1.25^{+0.07}_{-0.06}$ &$1.05^{+0.82}_{-0.64}$&$10.3^{+1.6}_{-1.3}$\cite{Hermosa:2017}&$0.34^{+0.01}_{-0.01}$\cite{Hermosa:2017}& 17\cite{Kirby:2017}&$\sim$-1.5\cite{Kirby:2017}\\
\hline

Draco  &$0.17^{+0.01}_{-0.01}$ & $0.56^{+0.02}_{-0.02}$ &$0.1^{+0.09}_{-0.05}$&$11^{+2.1}_{-1.5}$\cite{Massari:2020}&$0.23^{+0.01}_{-0.01}$
\cite{McConnachie:2020}&2.2\cite{Lokas:2005} &$\sim$-1.9\cite{Aparicio:2001}\\
\hline

Leo I  &$0.24^{+0.01}_{-0.01}$ & $1.30^{+0.08}_{-0.08}$ &$1.75^{+0.78}_{-0.96}$&$9.2^{+1.2}_{-1.2}$\cite{McConnachie:2020}&$0.26^{+0.01}_{-0.01}$\cite{Battaglia:2022}&$34^{+11}_{-11}$\cite{Koch:2007}&$\sim$-1.45 \cite{McConnachie:2020}\\
\hline
Phoenix  &$0.28^{+0.05}_{-0.06}$ & $1.27^{+0.01}_{-0.01}$ &$1.1^{+0.54}_{-0.55}$&$9.3^{+0.7}_{-0.7}$\cite{McConnachie:2020}&$0.29^{+0.01}_{-0.01}$\cite{McConnachie:2020}&6.2\cite{Held:1999}&$\sim$-1.5\cite{McConnachie:2020}\\
\hline
Canes Ventici  &$0.308^{+0.019}_{-0.018}$ & $1.06^{+0.11}_{-0.10}$ &$2.29^{+1.77}_{-1.35}$&$7.6^{+0.4}_{-0.4}$\cite{McConnachie:2020}&$0.47^{+0.02}_{-0.02}$\cite{McConnachie:2020}&2.3\cite{Matus:2020}&$-1.98^{+0.01}_{-0.01}$\cite{McConnachie:2020}\\
\hline
Sextans  &$0.48^{+0.01}_{-0.01}$ & $1.31^{+0.05}_{-0.06}$ &$1.61^{+0.51}_{-0.49}$&$7.9^{+1.3}_{-1.3}$\cite{McConnachie:2020}&$0.715^{+0.01}_{-0.01}$\cite{Okamoto:2017}&$4.37^{+1.69}_{-1.69}$\cite{Battaglia:2011} & $\sim$-1.95\cite{McConnachie:2020}\\
\hline

Crater II  &$0.71^{+0.09}_{-0.08}$ & $1.68^{+0.25}_{-0.23}$ &$2.6^{+}_{-}$&$2.7^{+0.3}_{-0.3}$\cite{Caldwell:2017}&$1.066^{+0.084}_{-0.084}$\cite{Caldwell:2017}&0.83\cite{Caldwell:2017}&$-1.98^{+0.1}_{-0.1}$\cite{Caldwell:2017}\\
\hline

Leo II  & $0.17^{+0.01}_{-0.01}$& $0.66^{+0.02}_{-0.01}$ &$3.76^{+0.79}_{-1.07}$&$7.4^{+0.4}_{-0.4}$\cite{Battaglia:2022}&$0.191^{+0.02}_{-0.02}$\cite{Moskowitz:2020}&$7.4^{+2}_{-2}$\cite{Koch:2007b}&$\sim$-1.65\cite{McConnachie:2020} \\
\hline
Carina  &$0.21^{+0.01}_{-0.01}$ & $0.81^{+0.04}_{-0.04}$ &$1.17^{+0.51}_{-0.61}$&$6.6^{+1.2}_{-1.2}$\cite{McConnachie:2020}&$0.424^{+0.06}_{-0.04}$\cite{Hayashi:2018}&5.9\cite{deBoer:2014}& $-1.72^{+0.01}_{-0.01}$\cite{McConnachie:2020}\\
\hline
Ursa Minor  &$0.28^{+0.01}_{-0.01}$ & $0.96^{+0.05}_{-0.04}$ &$0.52^{+0.9}_{-0.4}$&$11.5^{+0.9}_{-0.8}$\cite{ Pace:2020}&$0.4675^{+0.06}_{-0.06}$\cite{Pace:2020}&3\cite{Carrera:2002}&$\sim$-2.13 \cite{McConnachie:2020}\\
\hline
Sculptor  &$0.21^{+0.01}_{-0.01}$ & $0.72^{+0.07}_{-0.07}$ &$0.12^{+0.25}_{-0.09}$&$10.1^{+0.3}_{-0.3}$\cite{Battaglia:2022}&$0.289^{+0.01}_{-0.01}$\cite{Battaglia:2022}&$20.3^{+7.9}_{-7.9}$\cite{Bettinelli:2019}&$\sim$-1.45\cite{McConnachie:2020} \\
\hline

And I  &$0.52^{+0.02}_{-0.02}$ & $1.47^{+0.05}_{-0.07}$ &$0.13^{+0.05}_{-0.02}$&$9.4^{+1.7}_{-1.5}$\cite{Kirby:2020}&$0.66^{+0.07}_{-0.07}$\cite{Collins:2013}&$23.98^{+0.57}_{-0.54}$\cite{Martin:2016}&$-1.51^{+0.02}_{-0.02}$ \cite{Kirby:2020}\\
\hline

And III  &$0.33^{+0.02}_{-0.02}$ & $1.38^{+0.12}_{-0.11}$ &$2.53^{+1.56}_{-0.89}$&$11.0^{+1.9}_{-1.6}$\cite{Kirby:2020}&$0.41^{+0.04}_{-0.04}$\cite{Martin:2016}&$4.78^{+0.11}_{-0.11}$\cite{Martin:2016}&$-1.75^{+0.01}_{-0.01}$ \cite{Kirby:2020}\\
\hline

And V  &$0.25^{+0.01}_{-0.01}$ & $0.81^{+0.05}_{-0.06}$ &$2.59^{+1.47}_{-1.37}$&$11.2^{+1.1}_{-1.0}$\cite{Kirby:2020}&$0.35^{+0.04}_{-0.04}$\cite{Martin:2016}&$4.07^{+0.10}_{-0.09}$\cite{Martin:2016}&$-1.84^{+0.03}_{-0.03}$ \cite{Kirby:2020}\\
\hline

And IX  &$0.22^{+0.02}_{-0.02}$ & $0.70^{+0.08}_{-0.08}$ &$3.22^{+1.15}_{-1.46}$&$10.9^{+2.0}_{-2.0}$\cite{Alexander:2017}&$0.36^{+0.06}_{-0.05}$\cite{Martin:2016}&$1.99^{+0.52}_{-0.41}$\cite{Alexander:2017}&$-1.90^{+0.60}_{-0.60}$ \cite{Collins:2013}\\
\hline

And XIV &$0.33^{+0.02}_{-0.02}$ & $0.98^{+0.11}_{-0.12}$ &$1.77^{+1.83}_{-1.10}$&$5.4^{+1.3}_{-1.3}$\cite{Jason:2010}&$0.39^{+0.19}_{-0.20}$\cite{Collins:2013}&$1.99^{+0.52}_{-0.41}$\cite{Alexander:2017}&$\sim$ \cite{}\\
\hline

And XV &$0.19^{+0.02}_{-0.02}$ & $0.65^{+0.08}_{-0.09}$ &$1.88^{+1.81}_{-1.24}$&$11.0^{+7.0}_{-5.0}$\cite{McConnachie:2012}&$0.23^{+0.03}_{-0.02}$\cite{Collins:2013}&$1.25^{+0.74}_{-0.25}$\cite{Alexander:2017}&$\sim-1.1$ \cite{Collins:2013}\\
\hline

And XVIII &$0.25^{+0.03}_{-0.02}$ & $0.85^{+0.12}_{-0.09}$ &$2.99^{+1.27}_{-1.49}$&$9.7^{+2.3}_{-2.3}$\cite{McGaugh:2013}&$0.33^{+0.02}_{-0.02}$\cite{Collins:2013}&$3.98^{+2.32}_{-1.47}$\cite{McGaugh:2013}&$-1.80^{+0.50}_{-0.50}$ \cite{Collins:2013}\\
\hline

And XXI  &$0.51^{+0.06}_{-0.05}$ & $1.32^{+0.18}_{-0.15}$ &$3.16^{+}_{-}$&$6.1^{+1}_{-0.9}$\cite{Collins:2021}&$1.005^{+0.175}_{-0.175}$\cite{Collins:2021}&$3.2^{+0.8}_{-0.7}$\cite{Collins:2021}&$\sim$-1.8 \cite{Collins:2021}\\
\hline

And XXIII &$0.80^{+0.06}_{-0.06}$ & $2.20^{+0.18}_{-0.19}$ &$3.92^{+0.70}_{-1.35}$&$7.1^{+1.0}_{-1.0}$\cite{Alexander:2017}&$1.19^{+0.10}_{-0.10}$\cite{Collins:2013}&$6.30^{+1.64}_{-1.29}$\cite{Alexander:2017}&$-1.80^{+0.20}_{-0.20}$ \cite{Collins:2013}\\
\hline

And XXV &$0.42^{+0.05}_{-0.06}$ & $1.15^{+0.20}_{-0.18}$ &$2.79^{+1.30}_{-1.18}$&$3.0^{+1.2}_{-1.1}$\cite{Alexander:2017}&$0.55^{+0.10}_{-0.07}$\cite{Martin:2016}&$3.16^{+0.82}_{-0.65}$\cite{Alexander:2017}&$-1.80^{+0.50}_{-0.50}$\cite{Collins:2013}\\
\hline

\end{tabular}
\caption{Observations and $\psi$DM profile fits for Dwarf Spheroidal galaxies. Column 1: Dwarf galaxy name, Column 2: Core radius  $r_c$,  Column 3: Transition point $r_t$, Column 4: Stellar scale radius $r_{s*}$, Column 5: Observable projected velocity dispersion $\sigma_{los,obs}$, Column 6: Observable half-light radius $r_{half,obs}$, Column 7: Observable luminosity $L_{obs}$, Column 8: Observable metallicity. We have excluded centrally younger and more metal-rich stellar populations found in some of these dwarfs, which may be attributed to later gas infall. Instead, we adopted the metal-poor stellar and velocity dispersion profiles of Leo II, Carina, Ursa Minor, and Sculptor, with mean velocity dispersion values as follows: $7.96^{+1.39}_{-1.1}$\cite{ Spencer:2017}, $8.75^{+0.75}_{-0.75}$\cite{Wilkinson:2006,Fabrizio:2016}, $11.5^{+0.9}_{-0.8}$\cite{Pace:2020} and $10.7^{+1.4}_{-1.2}$\cite{Chen:2017}.
%with respective core radii of
%$r_{c,poor}$ (kpc): $0.18^{+0.02}_{-0.02}$, $0.29^{+0.02}_{-0.02}$, $0.35^{+0.04}_{-0.04}$ and %$0.27^{+0.01}_{-0.01}$. 
  }

\label{tabla:1}
\end{table*}

\newpage

\subsection{Ultra-Faint Dwarf Galaxies}\label{UFD}

Analysis of the core-halo structure has been conducted for all UFD stellar densities. Our predictions for the UFD class ($1.5 \times 10^{-21}$ eV) in $\psi$DM are illustrated in green, representing the 2$\sigma$ range of the posterior distribution of profiles. We have included nearly all the UFDs within the Local Group, incorporating stellar profile data points beyond 0.25 kpc. It is crucial to emphasize that all these galaxies are consistent with the core-halo structure, both in the Milky Way and Andromeda, underscoring the universality of this structural pattern for dwarf galaxies.\hfill\break

\begin{figure*}[h!]
  
  \centering
  \includegraphics[width=0.8\textwidth,height=15cm]{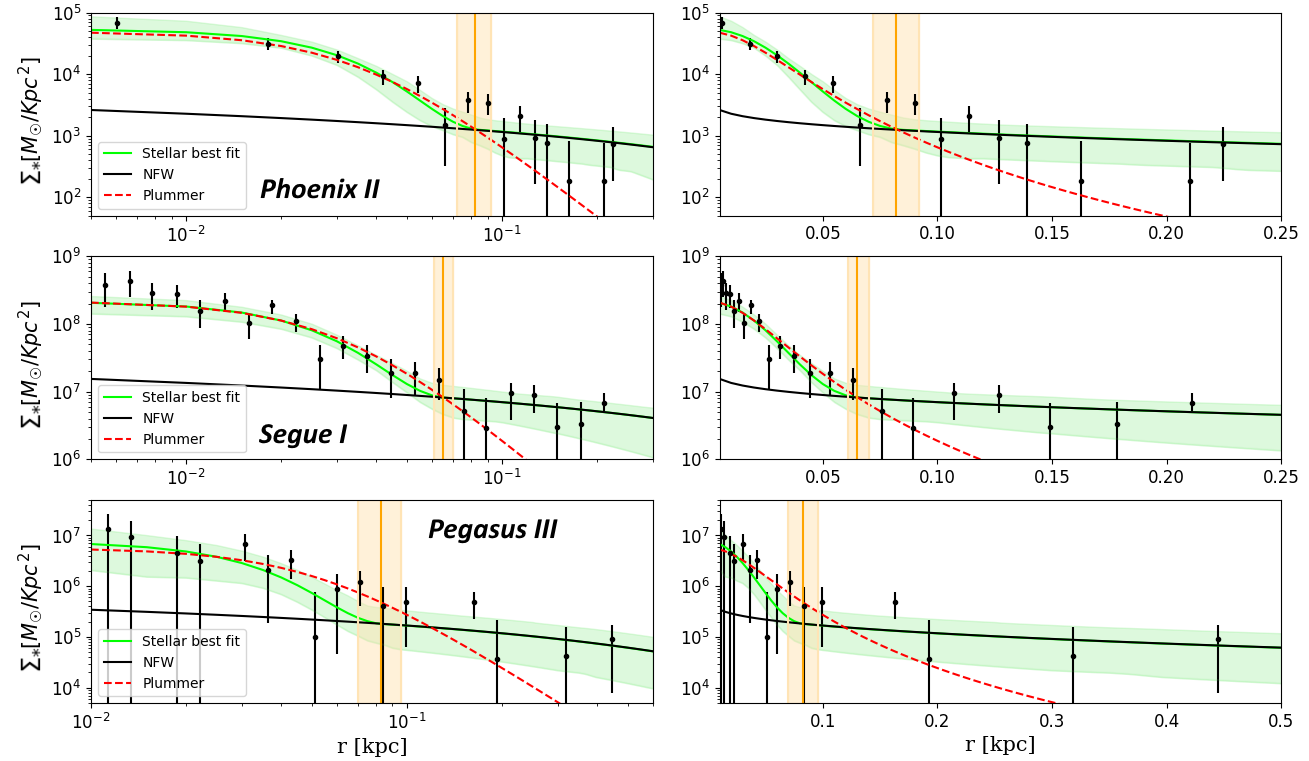}
  \caption{{\bf Ultra Faint Dwarfs:} This figure presents the star count profiles versus dwarf galaxy radius for the "ultra-faint" dwarf galaxies in the local group, as listed in Table \ref{tabla:2}. Many UFD dwarfs show clear evidence of extended halos stretching to approximately $\simeq 0.5$ kpc, most notably visible on the linear scale of the left-hand panel. Cores are also apparent on a scale of less than 0.1 kpc in these UFD dwarfs. A standard Plummer profile (red dashed curve) is observed to roughly fit the core region but falls significantly short at large radius. The soliton profile, normalized to the mean boson mass estimated for these dwarfs ($\simeq10^{-21}$ eV), is shown in green. The distinctive soliton profile provides an excellent fit to the observed cores, surrounded by a halo of excited states that azimuthally average to an approximately NFW-like profile beyond the soliton radius. The cores align well with the predicted form of the soliton profile, as best seen on a log scale in the right panels. The best-fit MCMC profile parameters are tabulated in the supplement. References for the data are as follows: Phoenix II \cite{Pakdil:2018}, Segue I \cite{Moskowitz:2020}, and Pegasus III \cite{Moskowitz:2020}.}\label{fig14}
\end{figure*}

\begin{figure*}[h!]
  
  \centering
  \includegraphics[width=0.8\textwidth,height=10cm]{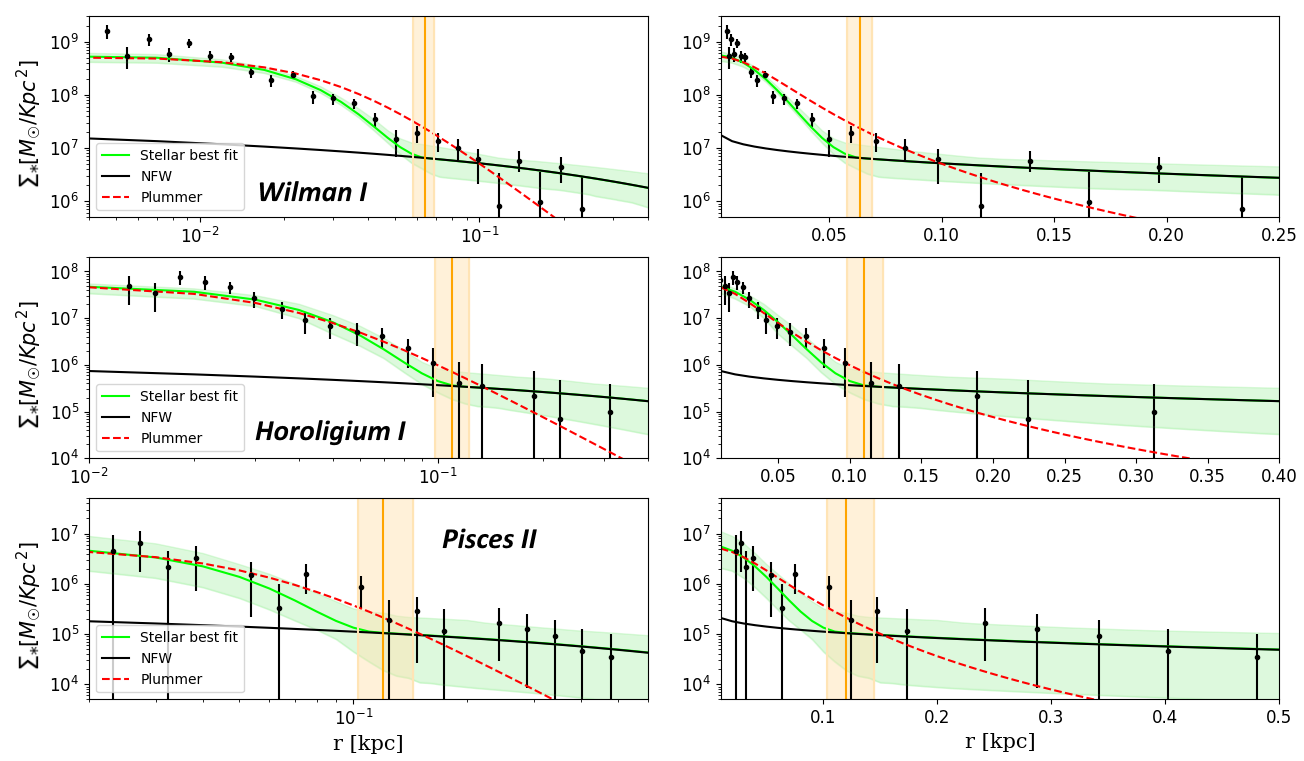}
  \caption{{\bf Ultra Faint Dwarf Galaxies:}  Similar to Figure \ref{fig11}, this figure includes three additional galaxies. References for the data are as follows: Wilman I \cite{Moskowitz:2020}, Horologium I \cite{Moskowitz:2020}, and Pisces II \cite{Moskowitz:2020}.}\label{fig15}
\end{figure*}

\begin{figure*}[h!]
  
  \centering
  \includegraphics[width=0.8\textwidth,height=10cm]{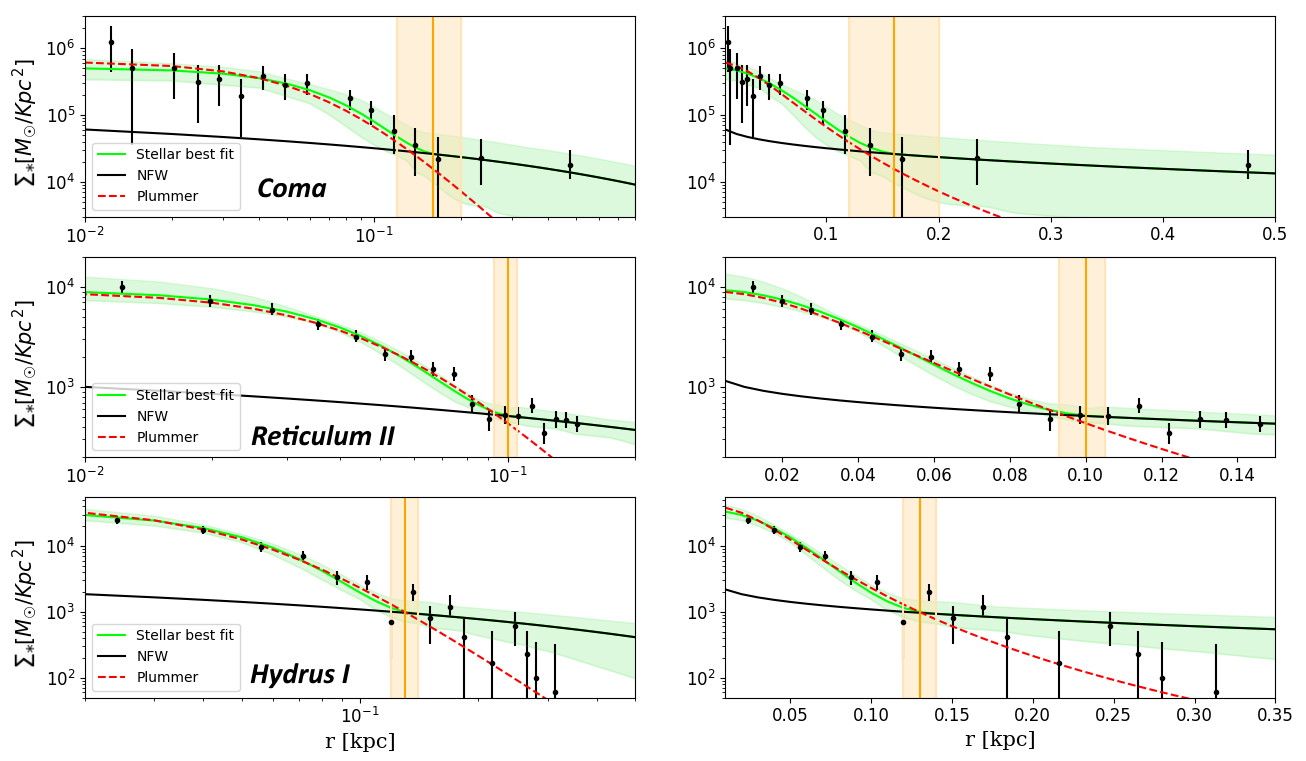}
  \caption{{\bf Ultra Faint Dwarf Galaxies:}  Similar to Figure \ref{fig11}, this figure includes three additional galaxies. References for the data are as follows: Coma Berenices \cite{Moskowitz:2020}, Reticulum II \cite{Koposov:2015}, and Hydrus I \cite{Koposov:2018}.}\label{fig16}
\end{figure*}

\begin{figure*}[h!]
  
  \centering
  \includegraphics[width=0.8\textwidth,height=9.75cm]{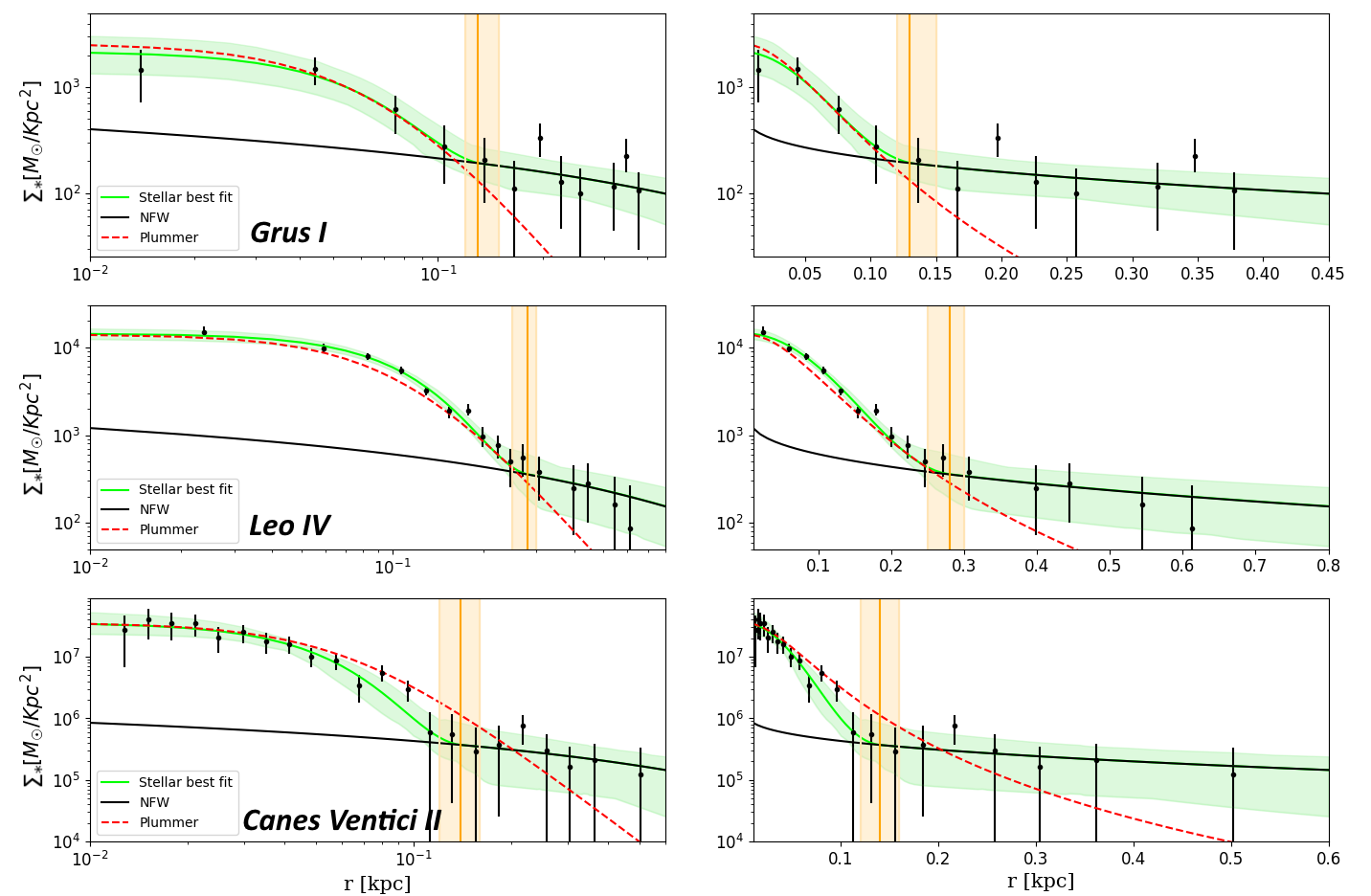}
  \caption{{\bf Ultra Faint Dwarf Galaxies:}  Similar to Figure \ref{fig11}, this figure includes three additional galaxies. References for the data are as follows: Grus I \cite{Koposov:2018}, Leo IV \cite{Okamoto:2012}, and Canes Venatici II \cite{Koposov:2018}.}\label{fig17}
\end{figure*}

\begin{figure*}[h!]
  
  \centering
  \includegraphics[width=0.8\textwidth,height=9.75cm]{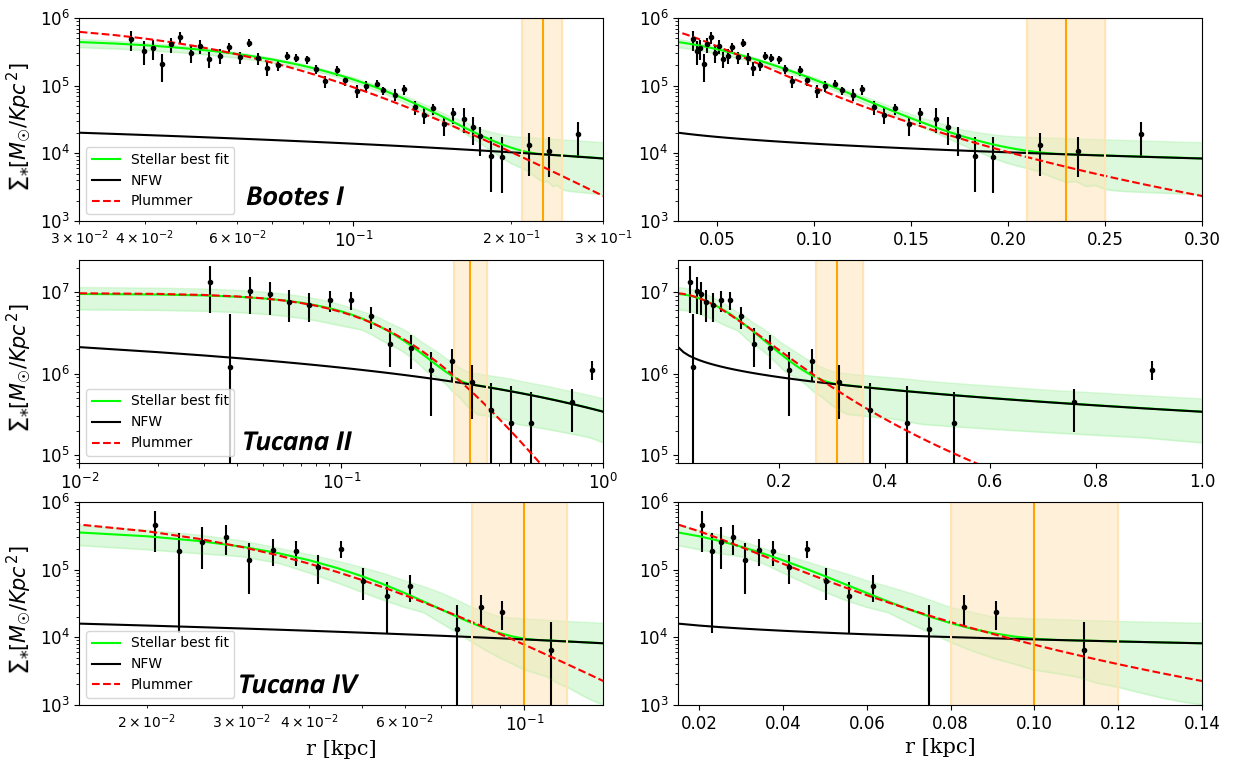}
  \caption{{\bf Ultra Faint Dwarf Galaxies:}  Similar to Figure \ref{fig11}, this figure includes three additional galaxies. References for the data are as follows: Bootes I \cite{Moskowitz:2020}, Tucana II \cite{Chiti:2022}, and Tucana IV \cite{Moskowitz:2020}. It's worth noting that Chiti et al. \cite{Chiti:2021} claimed a surprisingly extended halo of stars and dark matter, extending to 1 kpc in extent for Tucana II.}\label{fig17.5}
\end{figure*}

\begin{figure*}[h!]
  
  \centering
  \includegraphics[width=0.8\textwidth,height=10cm]{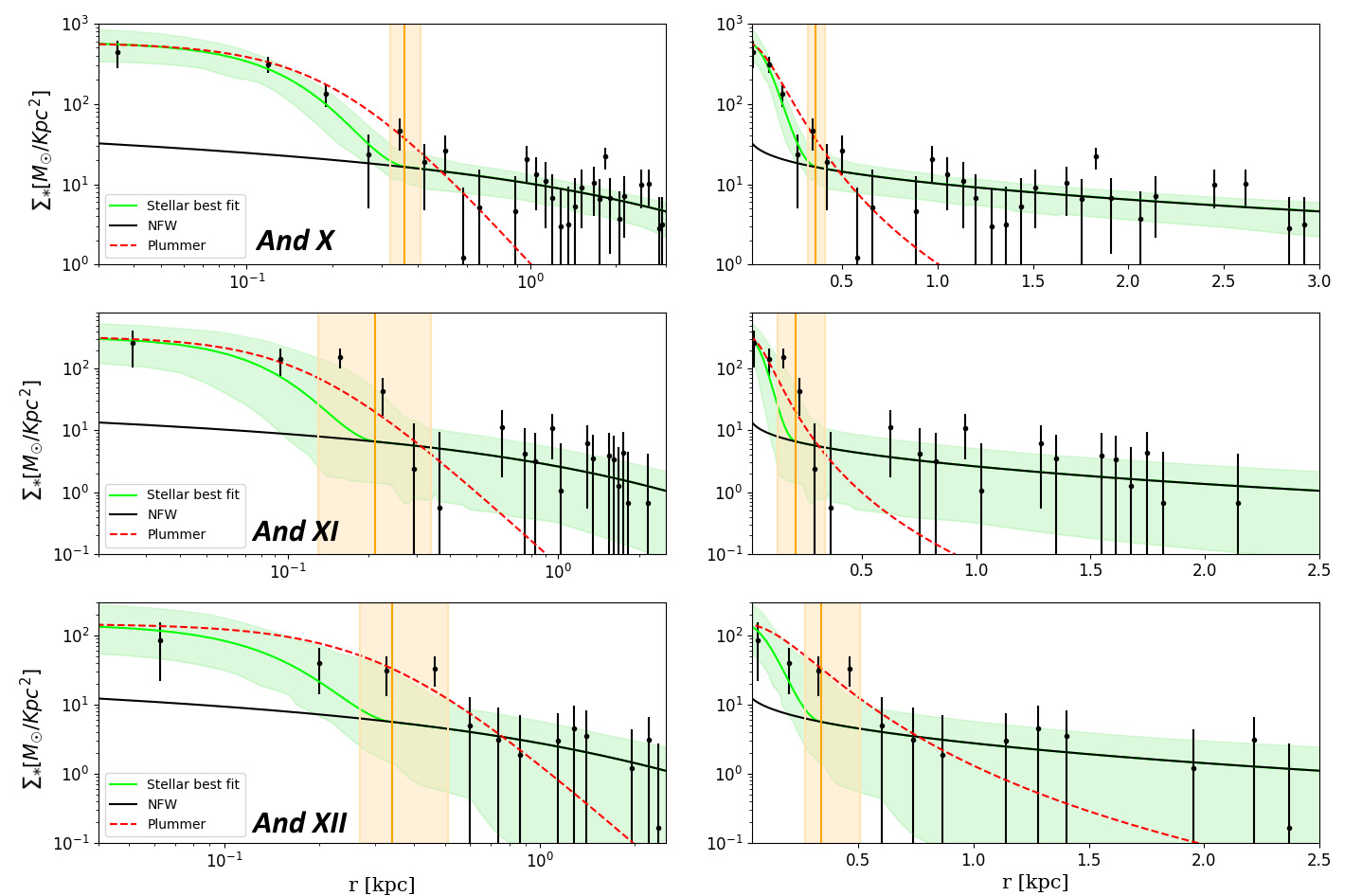}
  \caption{{\bf Ultra Faint Dwarf Galaxies:}  
Similar to Figure \ref{fig11}, this figure includes three additional galaxies. References for the data are as follows: Andromeda X \cite{Martin:2016}, Andromeda XI \cite{Martin:2016}, and Andromeda XII \cite{Martin:2016}.}\label{fig18}
\end{figure*}

\begin{figure*}[h!]
  
  \centering
  \includegraphics[width=0.8\textwidth,height=10cm]{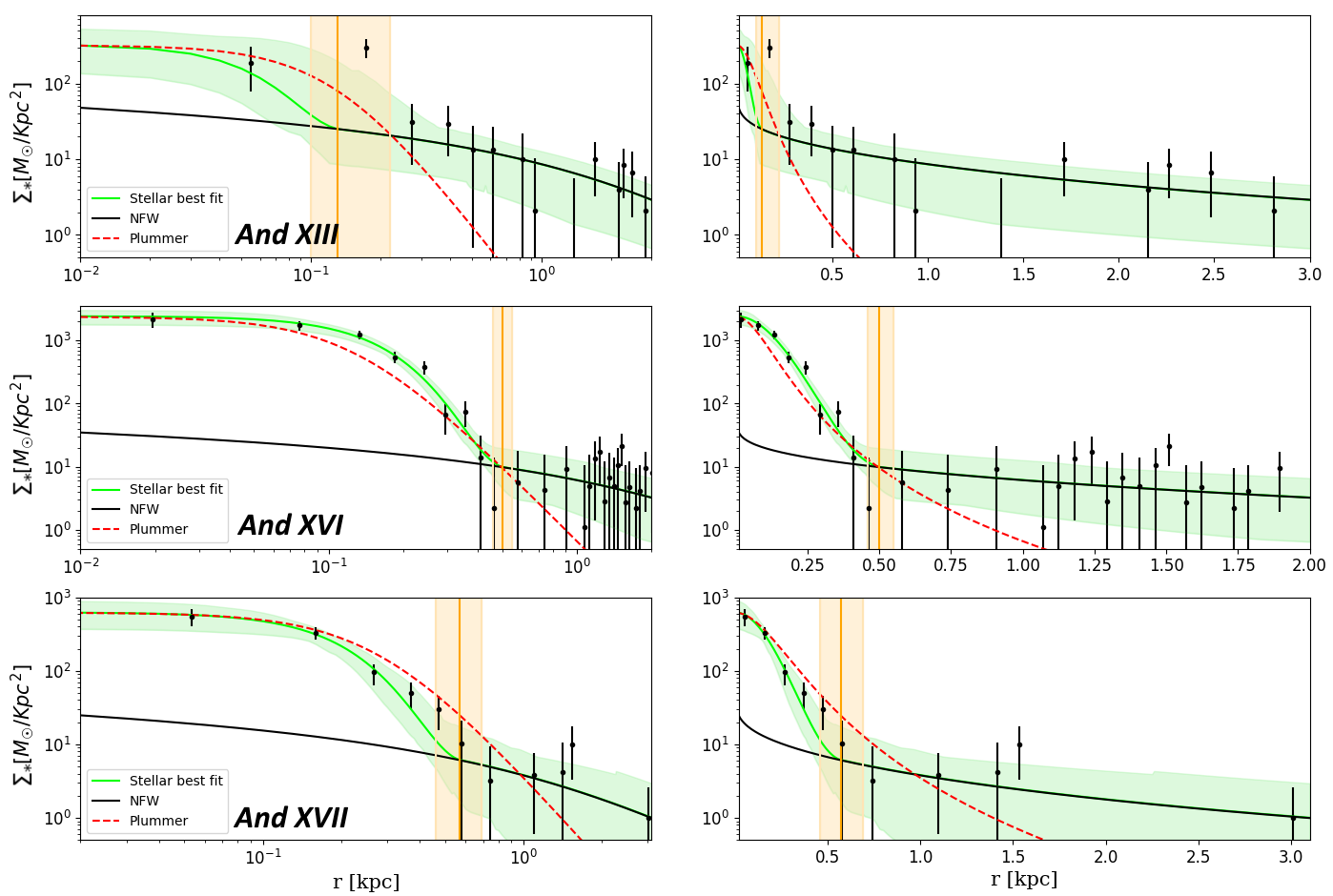}
  \caption{{\bf Ultra Faint Dwarf Galaxies:}  
Similar to Figure \ref{fig11}, this figure includes three additional galaxies. References for the data are as follows: Andromeda XIII \cite{Martin:2016}, Andromeda XVI \cite{Martin:2016}, and Andromeda XVII \cite{Irwin:2008}.}\label{fig19}
\end{figure*}

\begin{figure*}[h!]
  
  \centering
  \includegraphics[width=0.8\textwidth,height=8.5cm]{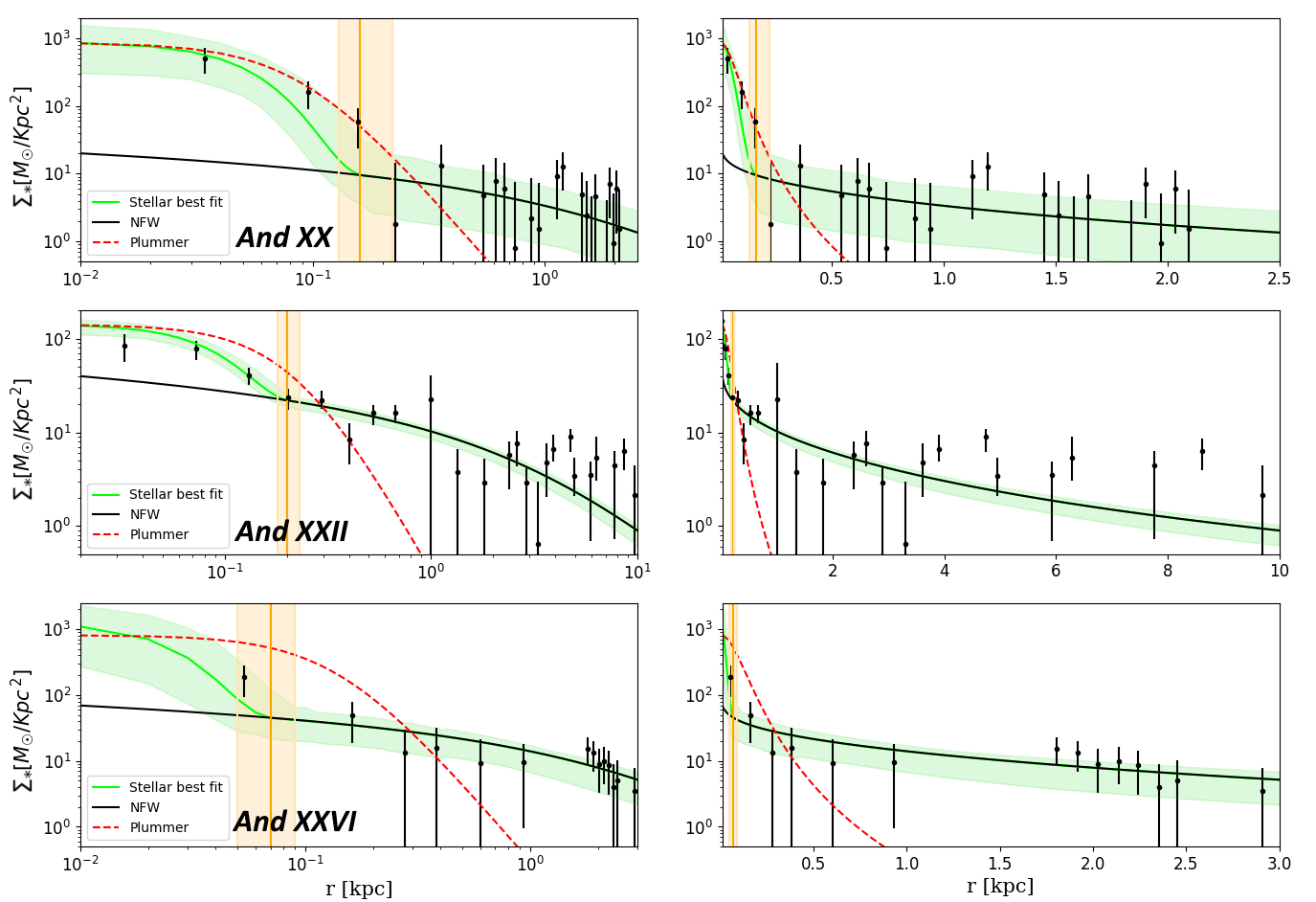}
  \caption{{\bf Ultra Faint Dwarf Galaxies:}  Similar to Figure \ref{fig11}, this figure includes three additional galaxies. References for the data are as follows: Andromeda XX \cite{Martin:2016}, Andromeda XXII \cite{Chapman:2013}, and Andromeda XXVI \cite{Martin:2016}.}\label{fig20}
\end{figure*}

\begin{table*}[h!]
	\centering

\begin{tabular}{|c|c|c|c|c|c|c|c|}

\hline

 Galaxy& $r_c$  & $r_{t}$& $r_{s*}$ &   $\sigma_{los}$ &$r_{half,obs}$& L,obs&  [Fe/H],obs \\
&(kpc)&  (kpc) &(kpc)  & (km/s)&(kpc)&$(10^{3}L_\odot)$&\\
\hline

Phoenix II  &$0.024^{+0.003}_{-0.003}$ &  $0.082^{+0.01}_{-0.01}$ &$1.37^{+1.17}_{-1.10}$ &$11^{+9.40}_{-5.3}$\cite{Battaglia:2022} &$0.036^{+0.008}_{-0.008}$\cite{Battaglia:2022}&$1.79^{+1.41}_{-0.79}$\cite{Munoz:2018} &$-2.51^{+0.19}_{-0.17}$\cite{Battaglia:2022}\\
\hline

Segue I  & $0.020^{+0.001}_{-0.001}$ &$0.065^{+0.005}_{-0.004}$&$1.59^{+0.81}_{-0.77}$& $3.9^{+0.8}_{-0.8}$\cite{McConnachie:2020}&$0.032^{+0.003}_{-0.003}$\cite{McConnachie:2020} &$0.28^{+0.27}_{-0.14}$\cite{Munoz:2018} & $-2.72^{+0.4}_{-0.4}$\cite{McConnachie:2020}\\
\hline

Pegasus III  &$0.024^{+0.004}_{-0.002}$ & $0.083^{+0.013}_{-0.013}$ &$1.16^{+1.10}_{-0.76}$&$5.4^{+3}_{-2.5}$\cite{Battaglia:2022}&$0.053^{+0.014}_{-0.014}$\cite{Battaglia:2022}&1.96\cite{Kim:2016} & $-2.55^{+0.15}_{-0.15}$\cite{Battaglia:2022}\\
\hline

Wilman I  & $0.0235^{+0.001}_{-0.001}$& $0.064^{+0.005}_{-0.006}$ &$0.64^{+1.21}_{-0.44}$&$4^{+0.8}_{-0.8}$\cite{Battaglia:2022}&$0.033^{+0.008}_{-0.008}$\cite{Battaglia:2022}&$0.87^{+0.86}_{-0.43}$\cite{Munoz:2018} & $\sim$-2.1\cite{Battaglia:2022}\\
\hline
%0.018 Rc wilman
Horoligium I  & $0.028^{+0.002}_{-0.002}$&$0.11^{+0.013}_{-0.012}$ &$1.29^{+1.01}_{-0.87}$ &$4.9^{+2.8}_{-0.9}$\cite{Battaglia:2022}&$0.041^{+0.01}_{-0.01}$\cite{Battaglia:2022}&$2.24^{+1.51}_{-0.90}$\cite{Munoz:2018}& $-2.76^{+0.10}_{-0.10}$\cite{Battaglia:2022}\\
\hline
Pisces II  &$0.032^{+0.006}_{-0.003}$ & $0.12^{+0.024}_{-0.017}$ &$1.85^{+1.30}_{-1.19}$&$5.4^{+3.6}_{-2.4}$\cite{Battaglia:2022}&$0.062^{+0.01}_{-0.01}$\cite{Battaglia:2022}&$4.16^{+1.76}_{-1.22}$\cite{Munoz:2018} & $-2.45^{+0.07}_{-0.07}$\cite{Battaglia:2022}\\
\hline
Coma Berenices   &$0.053^{+0.010}_{-0.011}$ &$0.16^{+0.04}_{-0.04}$  &$1.67^{+0.79}_{-0.90}$&$4.6^{+0.8}_{-0.8}$\cite{Battaglia:2022}&$0.069^{+0.005}_{-0.005}$\cite{Battaglia:2022}&$4.81^{+1.24}_{-0.99}$ \cite{Munoz:2018}&$-2.25^{+0.05}_{-0.05}$\cite{Battaglia:2022}\\
\hline
Reticulum II  &$0.0333^{+0.002}_{-0.002}$ & $0.10^{+0.005}_{-0.007}$ &$1.48^{+1.04}_{-0.85}$&$3.22^{+1.64}_{-0.49}$\cite{McConnachie:2020}&$0.053^{+0.002}_{-0.002}$\cite{McConnachie:2020}&$2.36^{+0.2}_{-0.2}$ \cite{Simon:2015}& $-2.65^{+0.07}_{-0.07}$\cite{Simon:2015}\\
\hline
Hydrus   &$0.041^{+0.003}_{-0.003}$ & $0.13^{+0.01}_{-0.01}$ &$1.27^{+1.01}_{-0.87}$&$2.69^{+0.51}_{-0.43}$\cite{McConnachie:2020}&$0.056^{+0.004}_{-0.004}$\cite{McConnachie:2020}&3.38\cite{Koposov:2018}  & $-2.52^{+0.09}_{-0.09}$\cite{Koposov:2018}\\
\hline

Grus I   &$0.0475^{+0.009}_{-0.008}$ & $0.13^{+0.02}_{-0.01}$ &$1.79^{+0.72}_{-0.76}$&$5.4^{+3}_{-2.5}$\cite{Battaglia:2022}&$0.070^{+0.025}_{-0.025}$\cite{McConnachie:2020}&$2.10^{+1.51}_{-0.88}$\cite{Munoz:2018} & $-1.88^{+0.09}_{-0.03}$\cite{Battaglia:2022}\\
\hline

Leo IV   & $0.084^{+0.003}_{-0.003}$& $0.28^{+0.02}_{-0.03}$ &$1.34^{+1.42}_{-0.34}$&$3.4^{+1.3}_{-0.9}$\cite{Jenkins:2021}&$0.114^{+0.01}_{-0.01}$\cite{Jenkins:2021}&$18^{+8}_{-8}$\cite{Blana:2012} & $-2.48^{+0.16}_{-0.13}$\cite{Jenkins:2021}\\
\hline

Canes Ventici II   & $0.037^{+0.005}_{-0.005}$& $0.14^{+0.02}_{-0.02}$ &$1.39^{+0.99}_{-0.93}$&$4.6^{+1.0}_{-1.0}$\cite{Battaglia:2022}&$0.07^{+0.01}_{-0.01}$\cite{Munoz:2018}&$10.46^{+3.05}_{-3.05}$\cite{Munoz:2018} & $-2.21^{+0.05}_{-0.05}$\cite{McConnachie:2020}\\
\hline

Bootes I   & $0.065^{+0.002}_{-0.003}$& $0.23^{+0.02}_{-0.02}$ &$1.86^{+0.73}_{-1.03}$&$2.4^{+0.9}_{-0.5}$\cite{Battaglia:2022}&$0.22^{+0.01}_{-0.01}$\cite{Battaglia:2022}&$21.78^{+5.64}_{-4.48}$\cite{Munoz:2018} & $-2.34^{+0.05}_{-0.05}$\cite{Jenkins:2021}\\
\hline

Tucana II   & $0.11^{+0.01}_{-0.01}$& $0.31^{+0.05}_{-0.04}$ &$2.45^{+0.98}_{-1.09}$&$2.8^{+1.2}_{-0.7}$\cite{Chiti:2022}&$0.12^{+0.03}_{-0.03}$\cite{Chiti:2022}&$\sim$2.83\cite{Munoz:2018} & $\sim$-2.7\cite{Chiti:2021}\\
\hline

Tucana IV   & $0.030^{+0.004}_{-0.003}$& $0.10^{+0.02}_{-0.02}$ &$1.90^{+1.31}_{-1.21}$&$4.3^{+1.7}_{-1.0}$\cite{Battaglia:2022}&$0.11^{+0.011}_{-0.009}$\cite{Moskowitz:2020}&$1.40^{+0.60}_{-0.30}$\cite{Simon:2020} & $-2.49^{+0.15}_{-0.16}$\cite{Battaglia:2022}\\
\hline

Leo V*   & $0.021^{+0.002}_{-0.001}$& $0.076^{+0.007}_{-0.007}$ &$1.61^{+1.50}_{-1.16}$&$3.7^{+2.3}_{-1.4}$\cite{McConnachie:2020}&$0.055^{+0.02}_{-0.02}$\cite{Battaglia:2022}&$4.92^{+1.93}_{-1.39}$\cite{Munoz:2018} & $-2.28^{+0.15}_{-0.16}$\cite{Battaglia:2022}\\
\hline

And X   & $0.10^{+0.01}_{-0.01}$& $0.36^{+0.05}_{-0.04}$ &$6.18^{+2.47}_{-2.87}$&$3.9^{+1.2}_{-1.2}$\cite{McGaugh:20132}&$0.21^{+0.04}_{-0.07}$\cite{Martin:2016}&$79.43^{+20.57}_{-16.}$\cite{Martin:2016} & $-2.27^{+0.03}_{-0.03}$\cite{Collins:2013}\\
\hline

And XI   & $0.059^{+0.03}_{-0.02}$& $0.21^{+0.13}_{-0.08}$ &$2.67^{+1.43}_{-1.44}$&$\leq4.6$\cite{Putman:2021}&$0.12^{+0.05}_{-0.04}$\cite{Martin:2016}&$25.12^{+14.69}_{-9.28}$\cite{Martin:2016} & $-2.0^{+0.20}_{-0.20}$\cite{Collins:2013}\\
\hline

And XII   & $0.10^{+0.05}_{-0.02}$& $0.34^{+0.17}_{-0.07}$ &$2.62^{+1.37}_{-1.33}$&$2.6^{+5.1}_{-2.6}$\cite{McGaugh:20132}&$0.32^{+0.06}_{-0.07}$\cite{Collins:2013}&$50.12^{+29.31}_{-18.50}$\cite{Martin:2016} & $-2.0^{+0.2}_{-0.2}$\cite{Collins:2013}\\
\hline

And XIII   & $0.045^{+0.03}_{-0.01}$& $0.13^{+0.09}_{-0.03}$ &$2.96^{+1.21}_{-1.28}$&$5.8^{+2.0}_{-2.0}$\cite{McGaugh:20132}&$0.13^{+0.08}_{-0.06}$\cite{Martin:2016}&$31.62^{+18.50}_{-15.77}$\cite{Martin:2016} & $-2.0^{+0.16}_{-0.13}$\cite{Collins:2013}\\
\hline

And XVI   & $0.12^{+0.01}_{-0.01}$& $0.50^{+0.05}_{-0.04}$ &$2.84^{+1.34}_{-1.57}$&$3.8^{+2.9}_{-2.9}$\cite{McGaugh:20132}&$0.13^{+0.03}_{-0.02}$\cite{Martin:2016}&$63.09^{+16.34}_{-12.97}$\cite{Martin:2016} & $-2.0^{+0.5}_{-0.5}$\cite{Collins:2013}\\
\hline

And XVII   & $0.15^{+0.02}_{-0.02}$& $0.57^{+0.12}_{-0.11}$ &$1.84^{+1.54}_{-1.25}$&$2.9^{+2.2}_{-1.9}$\cite{McGaugh:2013}&$0.29^{+0.06}_{-0.05}$\cite{Martin:2016}&$100.00^{+25.89}_{-20.57}$\cite{Martin:2016} & $\sim-2.0$\cite{Collins:2013}\\
\hline

And XX   & $0.042^{+0.01}_{-0.007}$& $0.16^{+0.06}_{-0.03}$ &$2.64^{+1.48}_{-1.27}$&$7.1^{+3.9}_{-2.5}$\cite{McGaugh:2013}&$0.09^{+0.04}_{-0.02}$\cite{Martin:2016}&$25.12^{+14.69}_{-9.27}$\cite{Martin:2016} & $-2.3^{+0.5}_{-0.5}$\cite{Collins:2013}\\
\hline

And XXII   & $0.078^{+0.01}_{-0.006}$& $0.20^{+0.03}_{-0.02}$ &$4.57^{+0.31}_{-0.47}$&$2.8^{+2.9}_{-1.4}$\cite{McGaugh:2013}&$0.23^{+0.08}_{-0.08}$\cite{Martin:2016}&$39.81^{+23.29}_{-19.86}$\cite{Martin:2016} & $-1.85^{+0.10}_{-0.10}$\cite{Collins:2013}\\
\hline

And XXVI   & $0.021^{+0.005}_{-0.004}$& $0.07^{+0.02}_{-0.02}$ &$3.65^{+0.86}_{-1.19}$&$8.6^{+2.8}_{-2.2}$\cite{McGaugh:2013}&$0.15^{+0.14}_{-0.08}$\cite{Martin:2016}&$15.85^{+24.00}_{-9.57}$\cite{Martin:2016} & $-1.9^{+0.20}_{-0.20}$\cite{Collins:2013}\\
\hline

\hline

\hline

\end{tabular}
\caption{Observations and $\psi$DM profile fits to ultra faint dwarf galaxies. Column 1: UFD name, Column 2: Core radius  $r_c$,  Column 3: Transition point $r_t$, Column 4: Stellar scale radius $r_{s*}$, Column 5: Observable projected velocity dispersion $\sigma_{los,obs}$, Column 6: Observable half-light radius $r_{half,obs}$, Column 7: Observable luminosity $L_{obs}$, Column 8: Observable age, Column 9: Observable metallicity. Note: Leo V has recently been suggested not to be a galaxy.}

\label{tabla:2}
\end{table*}

\newpage

\begin{figure*}[h!]
  
  \centering
  \includegraphics[width=1\textwidth,height=10.5cm]{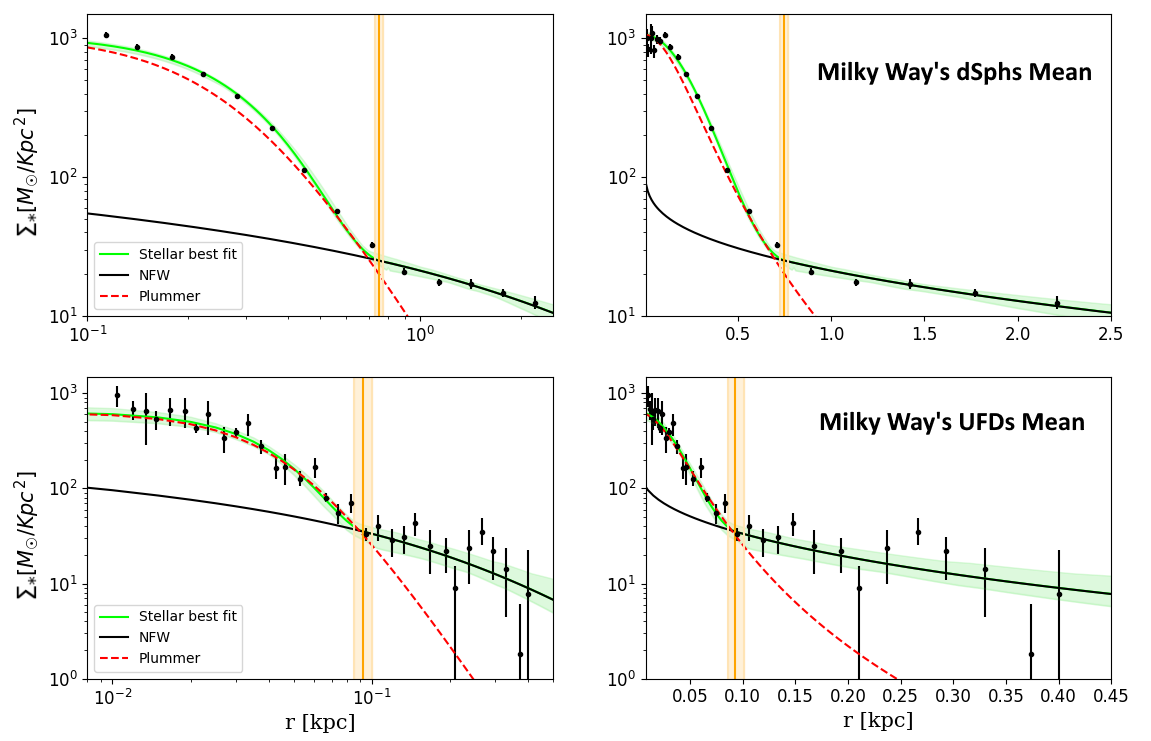}
  \caption{Like Figure \ref{figmeans} but just for Milky Way's satellites.}\label{figmeansmilky}
\end{figure*}

\begin{figure*}[h!]
  
  \centering
  \includegraphics[width=1\textwidth,height=10cm]{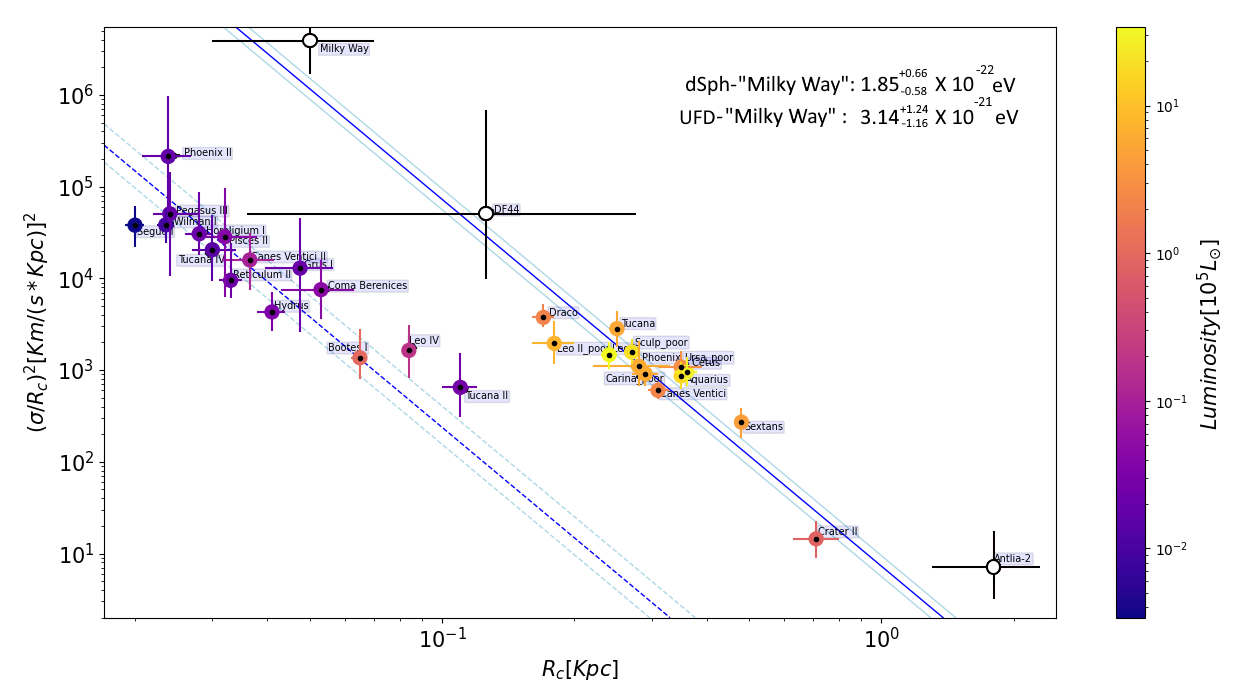}
  \caption{{\bf DM density vs Core radius}. Like Figure \ref{figrcl} just for Milky Way's satellites.}\label{figrclMilky}
\end{figure*}

\begin{figure*}[h!]
  
  \centering
  \includegraphics[width=1\textwidth,height=10cm]{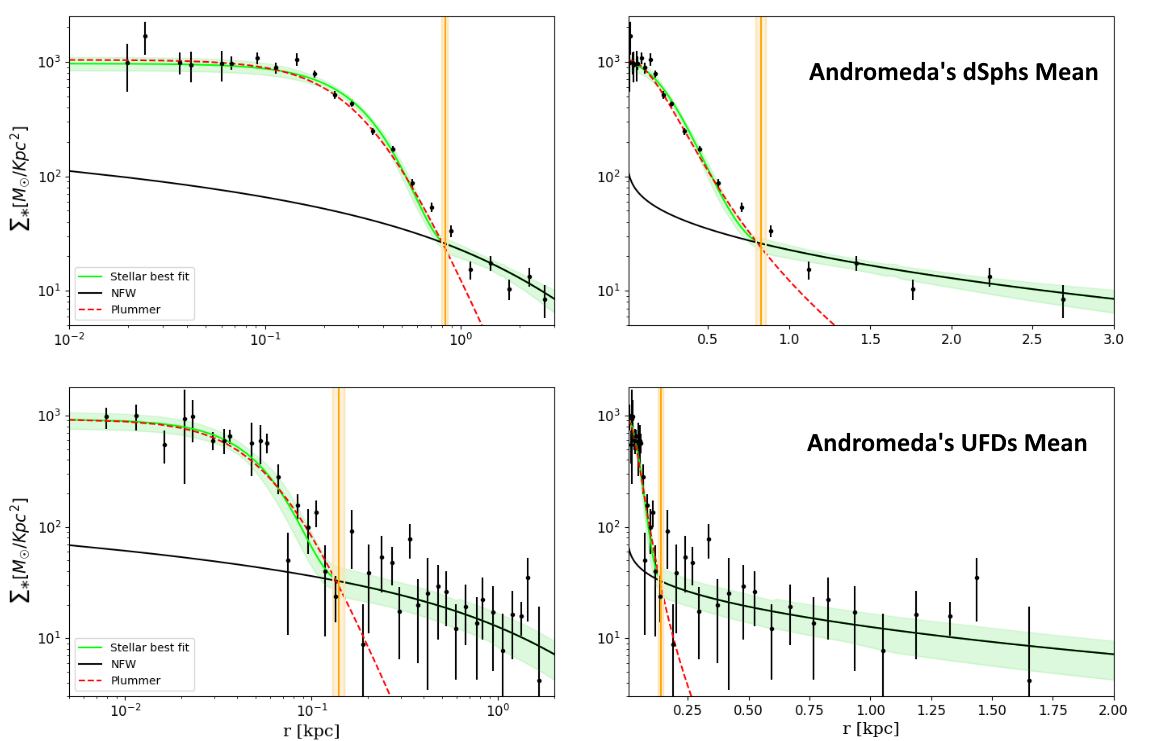}
  \caption{Like Figure \ref{figmeans} but just for Andromeda's satellites.}\label{figmeansandro}
\end{figure*}

\begin{figure*}[h!]
  
  \centering
  \includegraphics[width=1\textwidth,height=10cm]{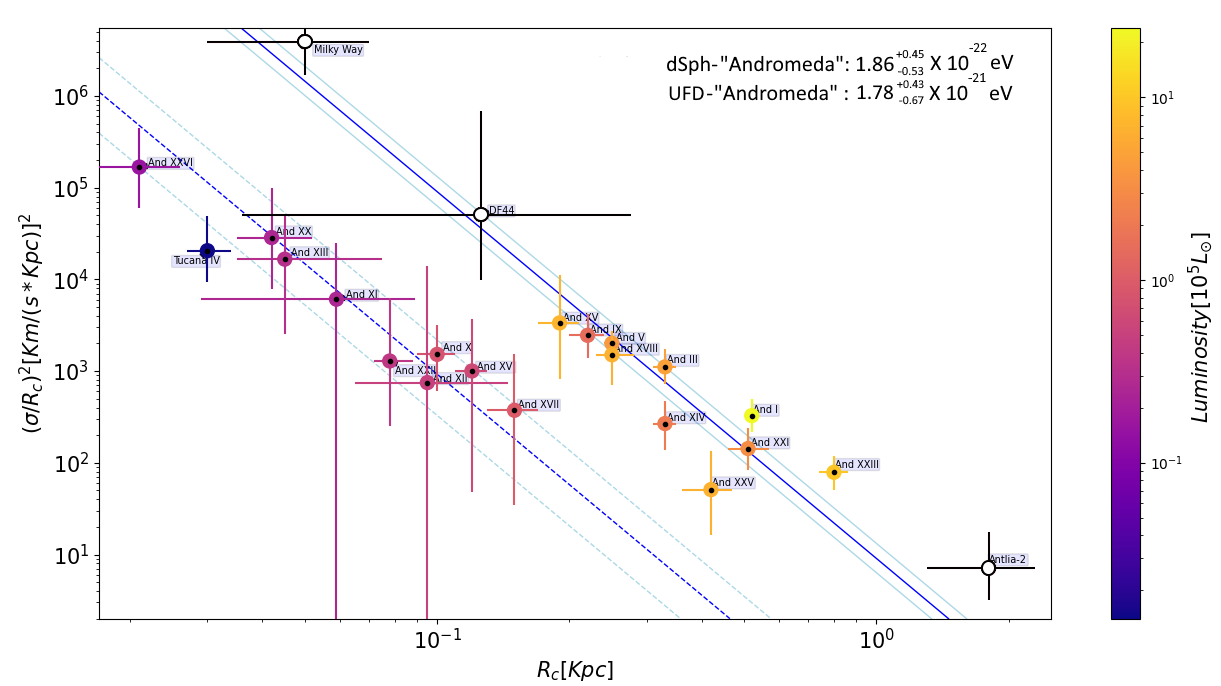}
  \caption{{\bf DM density vs Core radius}. Like Figure \ref{figrcl} but just for Andromeda's satellites.}\label{figrclandro}
\end{figure*}

In this analysis, we examine the satellites of the Milky Way and Andromeda separately to determine if there are any differences in core-halo structure, as depicted in Figures \ref{figmeansmilky} and \ref{figmeansandro}. Additionally, we explore the density versus core radius trend in Figures \ref{figrclMilky} and \ref{figrclandro}.

\subsection{Milky Way}

Milky way's galaxies alone.

\begin{figure*}[h!]
  
  \centering
  \includegraphics[width=0.7\textwidth,height=6.5cm]{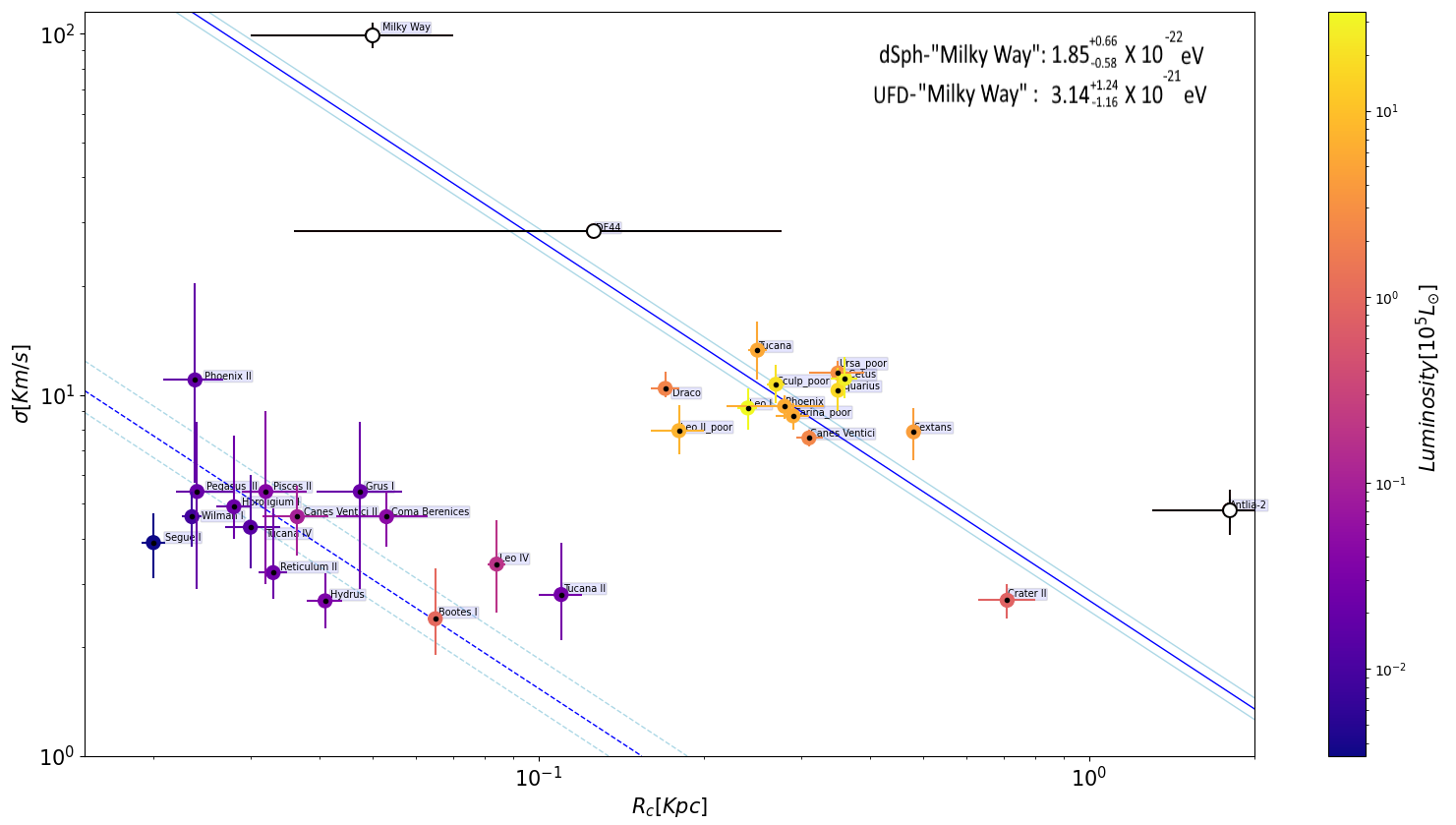}
  \caption{{\bf Velocity dispersion Vs Core radius}. Like Figure \ref{figrcd} but just for Milky Way's satellites.}\label{Milkysigma}
\end{figure*}

\begin{figure*}[h!]
  
  \centering
  \includegraphics[width=1\textwidth,height=10cm]{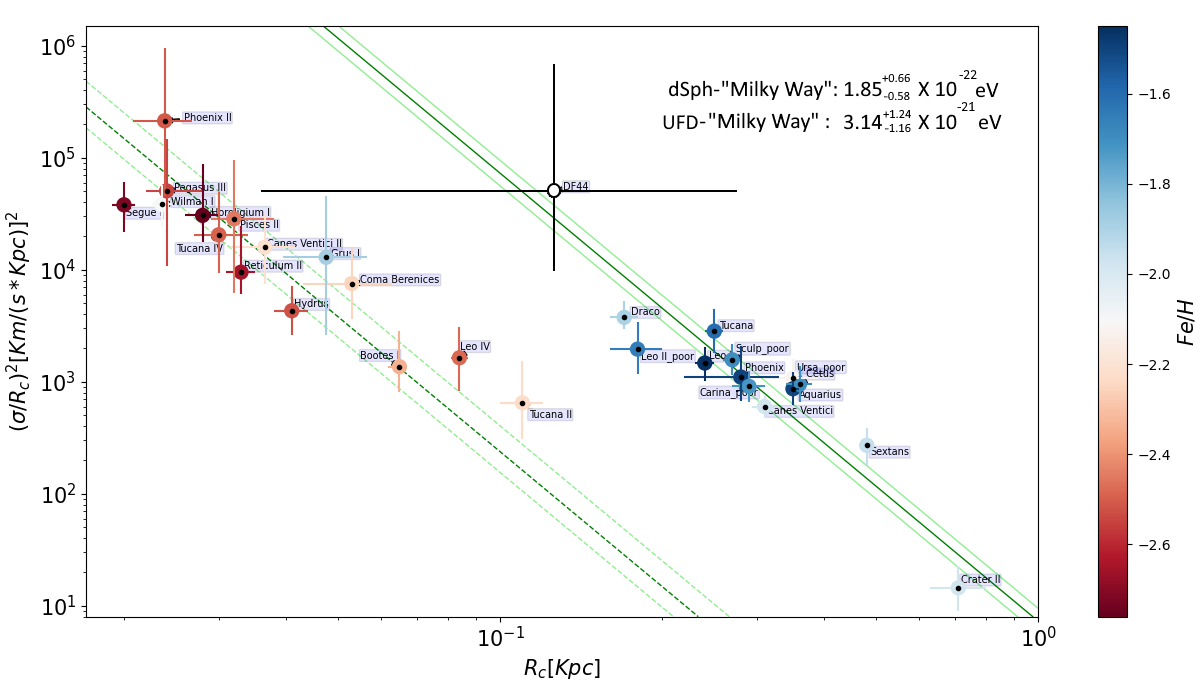}
  \caption{{\bf DM density vs Core radius}. Like Figure \ref{figrcl2} but just for Milky Way's satellites.}\label{Milkyrcfe}
\end{figure*}

\begin{figure*}[h!]
  
  \centering
  \includegraphics[width=0.8\textwidth,height=9cm]{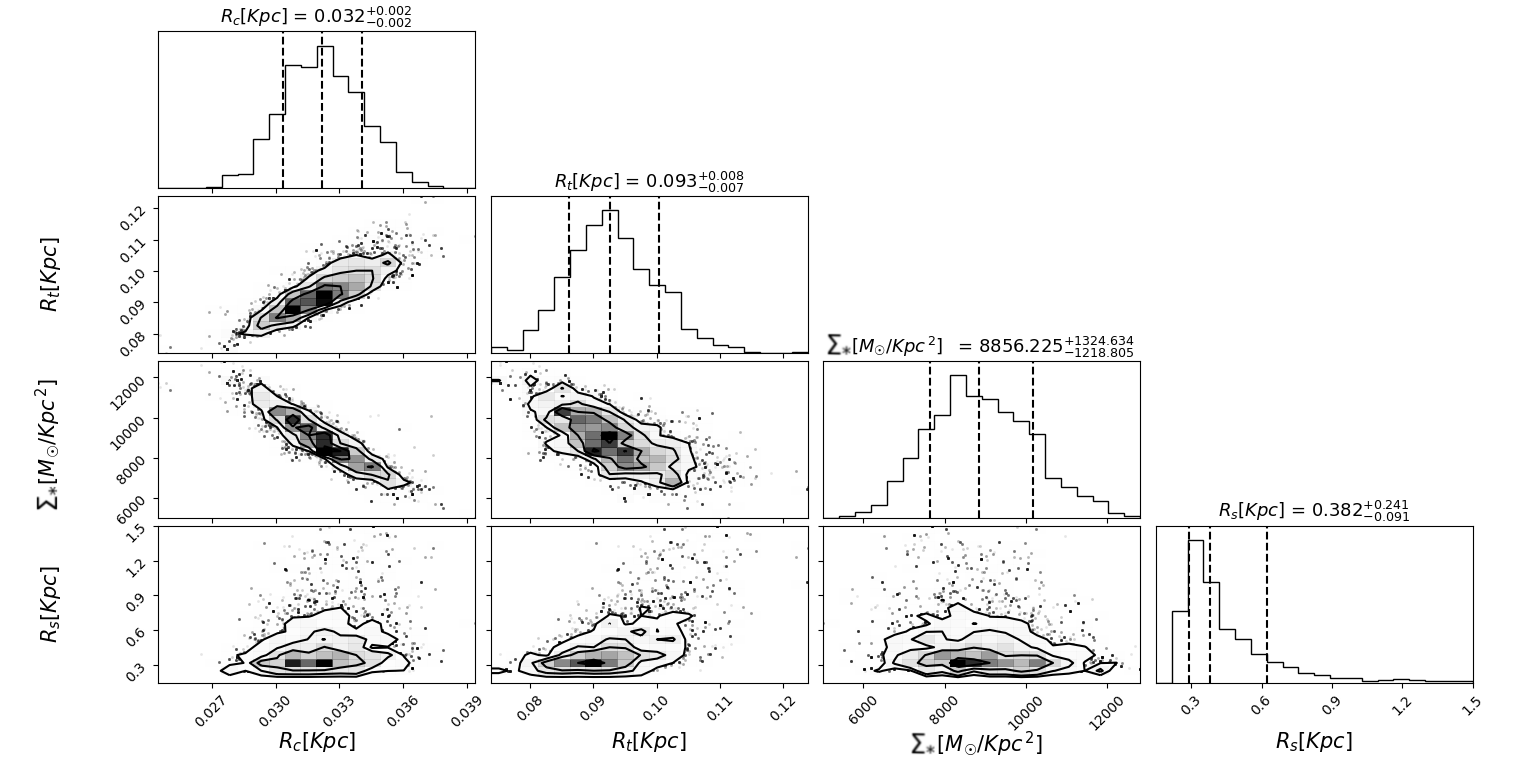}
  \caption{Milky's UFD  mean( Figure \ref{figmeansmilky} low panel): correlated distributions of the free parameters.  As can be seen the core radius and transition radius are well defined despite the wide Gaussian priors, indicating a reliable result. The contours represent the 68\%, 95\%, and 99\% confidence levels. The best-fit parameter values are the medians(with errors), represented by the dashed black ones, and tabulated in Table \ref{tabla:collage}.}\label{corner2mean}
\end{figure*}

\begin{figure*}[h!]
  
  \centering
  \includegraphics[width=0.8\textwidth,height=9cm]{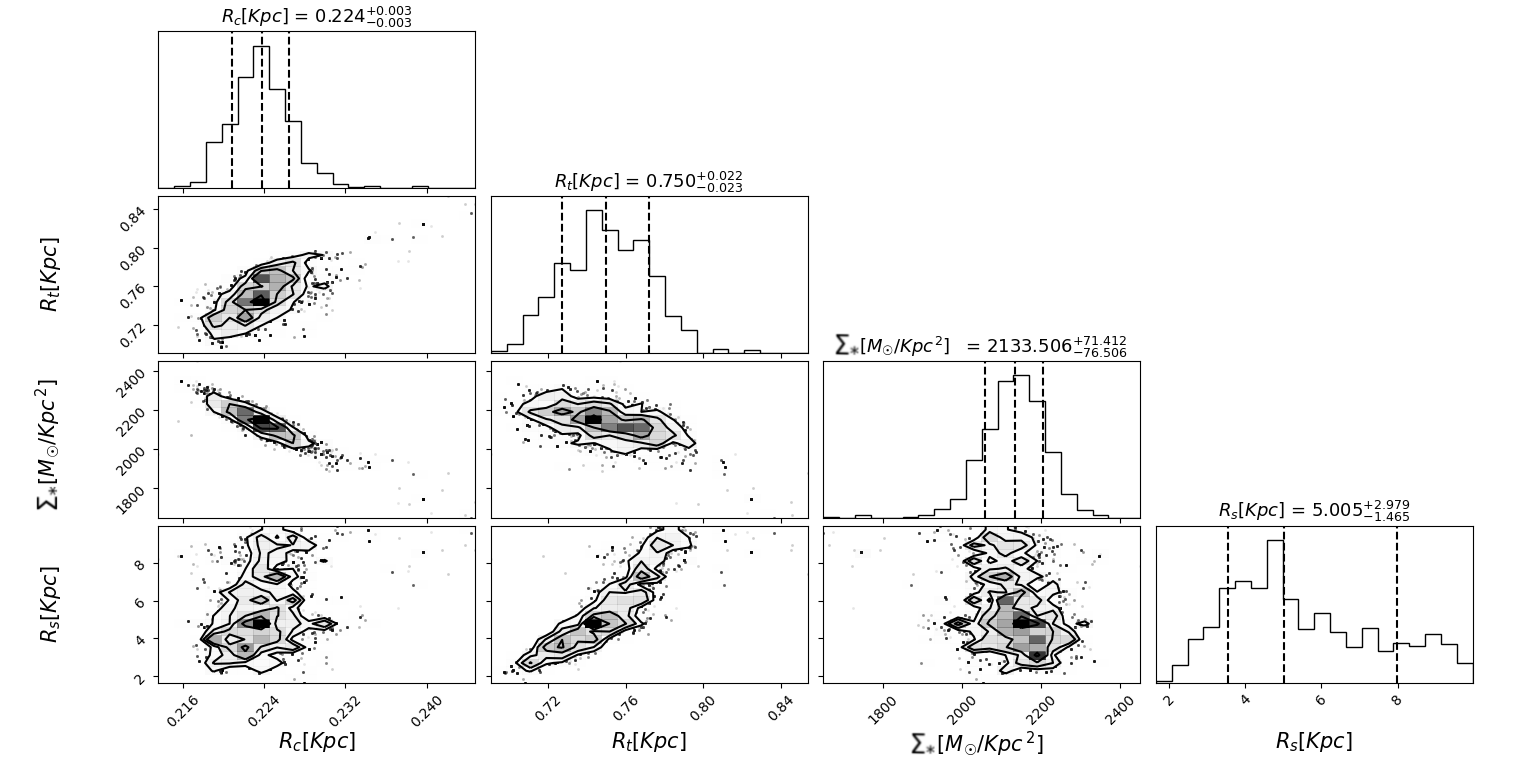}
  \caption{Milky's dSph mean( Figure \ref{figmeansmilky} top panel): correlated distributions of the free parameters.  As can be seen the core radius and transition radius are well defined despite the Gaussian input priors, indicating a reliable result. The contours represent the 68\%, 95\%, and 99\% confidence levels. The best-fit parameter values are the medians(with errors), represented by the dashed black ones, and tabulated in Table \ref{tabla:collage}.}\label{corner1mean}
\end{figure*}

\newpage
\vspace{10cm}
\subsection{Andromeda}

Andromeda's galaxies alone.

\begin{figure*}[h!]
  
  \centering
  \includegraphics[width=0.7\textwidth,height=6.5cm]{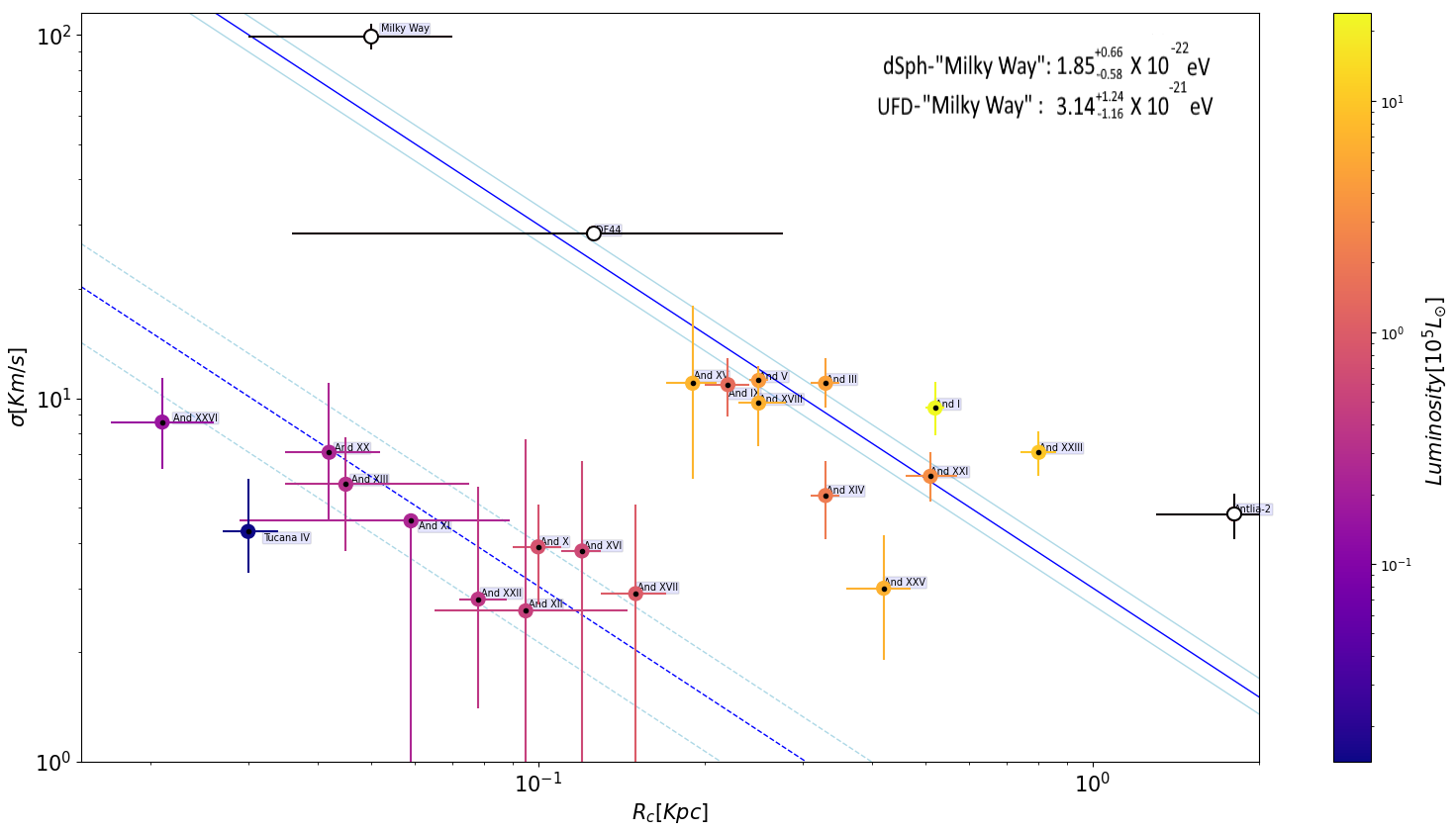}
  \caption{{\bf Velocity dispersion Vs Core radius}. Like Figure \ref{figrcd} but just for Andromeda's satellites.}\label{androsigma}
\end{figure*}

\begin{figure*}[h!]
  
  \centering
  \includegraphics[width=1\textwidth,height=10cm]{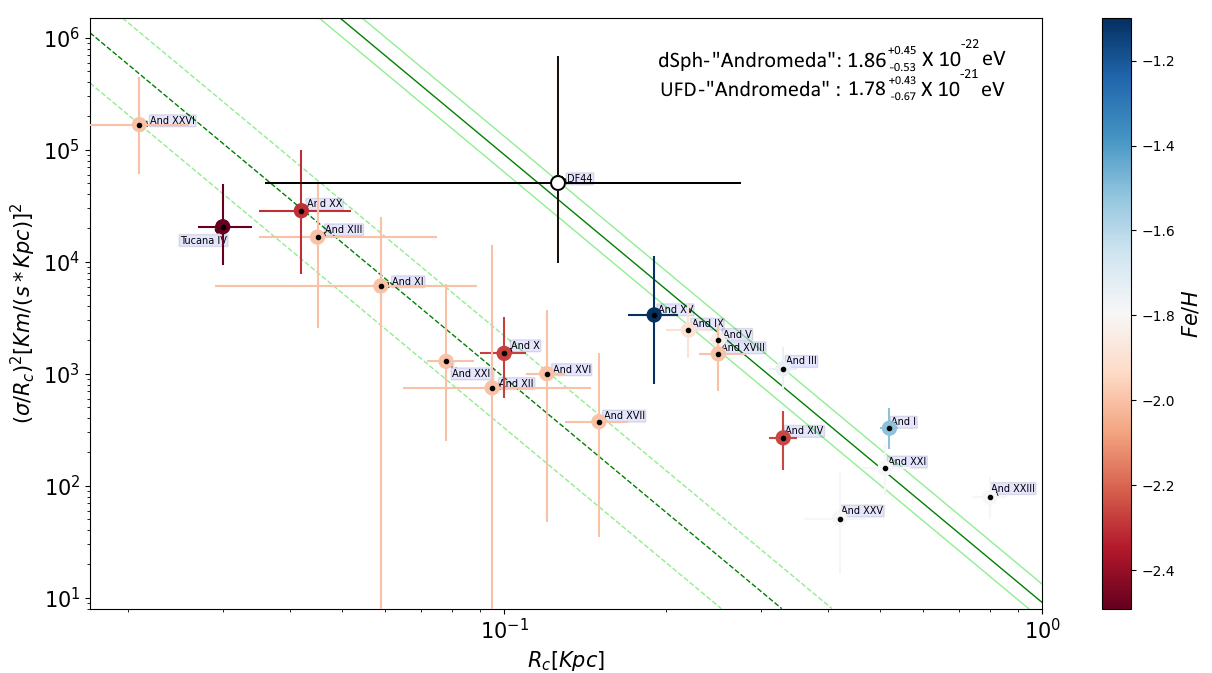}
  \caption{{\bf DM density vs Core radius}. Like Figure \ref{figrcl2} but just for Andromeda's satellites.}\label{androrcfe}
\end{figure*}

\begin{figure*}[h!]

  \centering
  \includegraphics[width=0.8\textwidth,height=9cm]{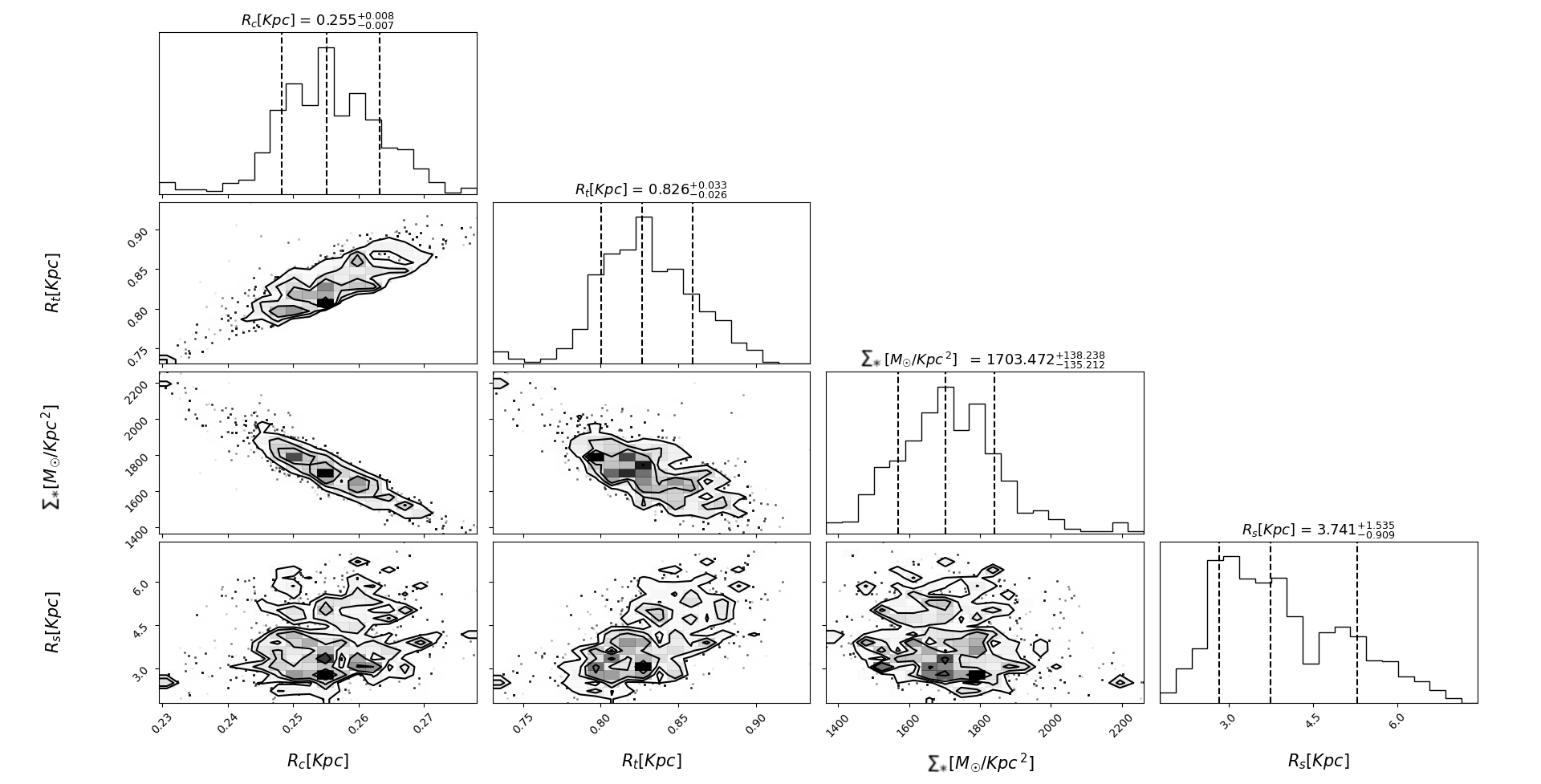}
  \caption{Andromeda's dSph  mean( Figure \ref{figmeansmilky} low panel): correlated distributions of the free parameters.  As can be seen the core radius and transition radius are well defined despite the Gaussian input priors, indicating a reliable result. The contours represent the 68\%, 95\%, and 99\% confidence levels. The best-fit parameter values are the medians(with errors), represented by the dashed black ones, and tabulated in Table \ref{tabla:collage}.}\label{corner1and}
\end{figure*}

\begin{figure*}[h!]

  \centering
  \includegraphics[width=0.8\textwidth,height=9cm]{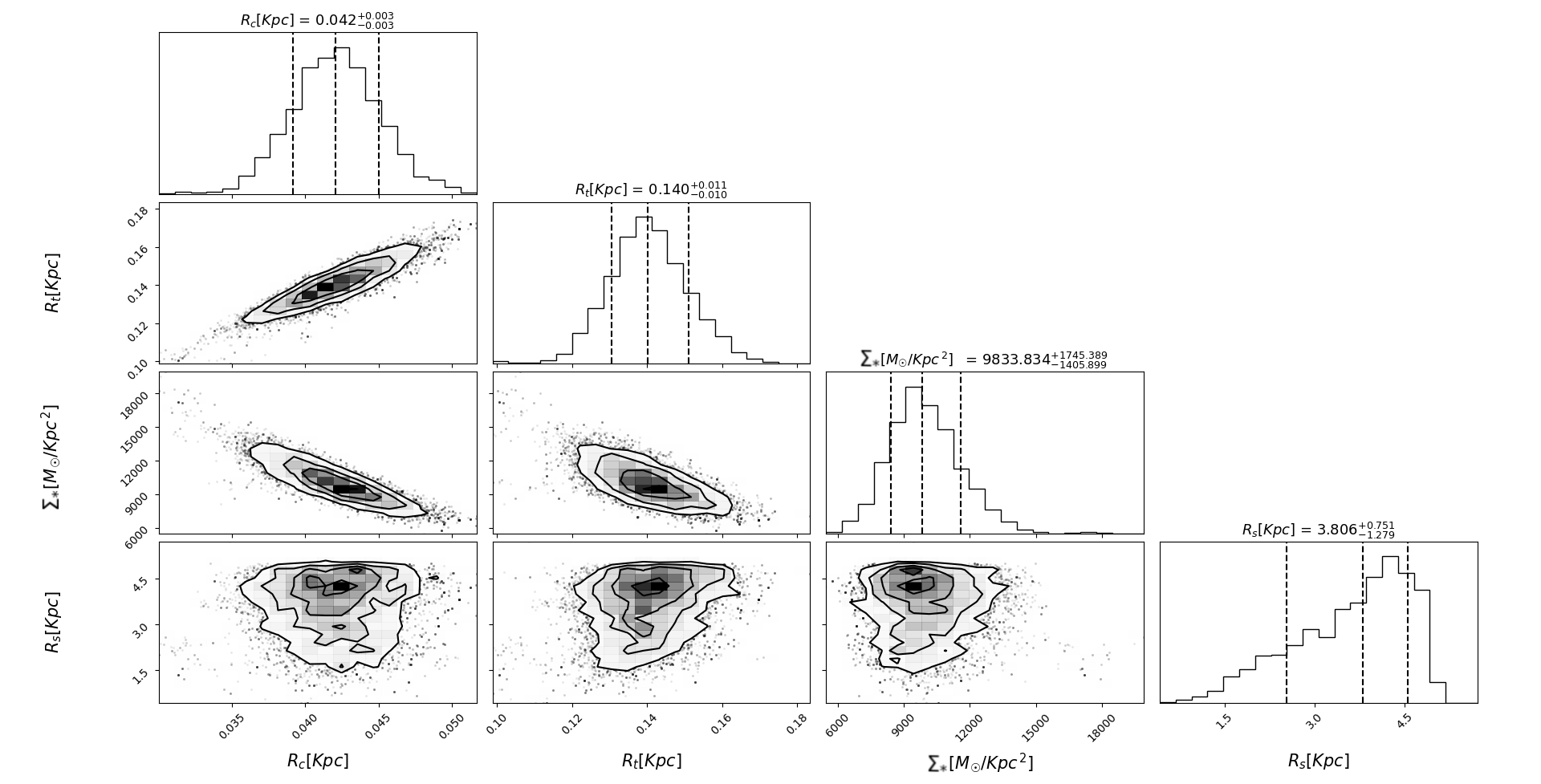}
  \caption{Andromeda's UFD  mean( Figure \ref{figmeansmilky} low panel): correlated distributions of the free parameters.  As can be seen the core radius and transition radius are well defined despite the Gaussian input priors, indicating a reliable result. The contours represent the 68\%, 95\%, and 99\% confidence levels. The best-fit parameter values are the medians(with errors), represented by the dashed black ones, and tabulated in Table \ref{tabla:collage}.}\label{corner2and}
\end{figure*}

\end{document}